\documentclass[11pt]{article}
\usepackage{amsmath,amssymb,amsfonts}
\usepackage[nosort]{cite}
\usepackage[margin=1in]{geometry}

\makeatletter \@addtoreset{equation}{section} \makeatother

\newcommand{\fft}[2]{{\frac{#1}{#2}}}
\newcommand{\ft}[2]{{\textstyle\frac{#1}{#2}}}

\def\nn{\nonumber}

\let\bm=\bibitem

\newcommand{\be}{\begin{equation}}
\newcommand{\ee}{\end{equation}}
\def\ba{\begin{array}}
\def\ea{\end{array}}
\def\ft#1#2{{\textstyle{\frac{\scriptstyle #1}{\scriptstyle #2}}}}
\def\fft#1#2{\frac{#1}{#2}}

\def\sst#1{{\scriptscriptstyle #1}}

\def\dalemb#1#2{{\vbox{\hrule height .#2pt
        \hbox{\vrule width.#2pt height#1pt \kern#1pt
                \vrule width.#2pt}
        \hrule height.#2pt}}}

\newcommand{\hoch}[1]{$\, ^{#1}$}
\newcommand{\bea}{\begin{eqnarray}}
\newcommand{\eea}{\end{eqnarray}}

\newcommand{\Tr}{\mbox{Tr}}
\def\0{{\sst{(0)}}}
\def\1{{\sst{(1)}}}
\def\2{{\sst{(2)}}}
\def\3{{\sst{(3)}}}
\def\4{{\sst{(4)}}}
\def\5{{\sst{(5)}}}
\def\6{{\sst{(6)}}}
\def\7{{\sst{(7)}}}
\def\8{{\sst{(8)}}}

\def\im{{{\rm i}}}

\def\R{\rlap{\rm I}\mkern3mu{\rm R}}

\def\R{\rlap{\rm I}\mkern3mu{\rm R}}

\def\R{{{\mathbb R}}}

\def\CP{{{\mathbb C}{\mathbb P}}}

\def\Z{{{\mathbb Z}}}

\begin{document}
\begin{flushright}
MCTP-07-13\ \ \ \ MIFP-07-14\\
{\bf arXiv:0705.2234}\\
May 2007
\end{flushright}

\vspace{10pt}
\begin{center}

{\Large {\bf New supersymmetric solutions of ${\cal N}=2$, $D=5$
gauged supergravity with hyperscalars}}

\vspace{20pt}

James T. Liu$^{\dagger}$, H. L\"u$^{\ddagger}$, C.N. Pope$^{\ddagger}$
and Justin F. V\'azquez-Poritz$^{\ddagger}$

\vspace{20pt}

{\hoch{\dagger}\it Michigan Center for Theoretical Physics\\
Randall Laboratory of Physics, The University of Michigan\\
Ann Arbor, MI 48109--1040, USA}

\vspace{10pt}

{\hoch{\ddagger}\it George P. \&  Cynthia W. Mitchell Institute
for Fundamental Physics\\
Texas A\&M University, College Station, TX 77843--4242, USA}

\vspace{40pt}

\underline{ABSTRACT}
\end{center}

We construct new supersymmetric solutions, including AdS bubbles, in
an $\mathcal N=2$ truncation of five-dimensional $\mathcal N=8$ gauged
supergravity.  This particular truncation is given by $\mathcal N=2$
gauged supergravity coupled to two vector multiples and three
incomplete hypermultiplets, and was originally investigated in the
context of obtaining regular AdS bubble geometries with multiple
active $R$-charges.  We focus on cohomogeneity-one solutions
corresponding to objects with two equal angular momenta and up to
three independent $R$-charges.  Curiously, we find a new set of zero
and negative mass solitons asymptotic to AdS$_5/\mathbb Z_k$, for
$k\ge3$, which are everywhere regular without closed timelike curves.

\newpage
\tableofcontents
\addtocontents{toc}{\protect\setcounter{tocdepth}{3}}
\newpage

\section{Introduction}

While sphere compactifications of string theory and M-theory have been
known for many years, they have taken on renewed importance since the
advent of the AdS/CFT correspondence \cite{agmoo}.  In particular, an
extremely well studied system is that of type IIB string theory on
AdS$_5\times S^5$ and the dual $\mathcal N=4$ super-Yang Mills gauge
theory.  In the supergravity limit, this system can be investigated
from both ten-dimensional and five-dimensional perspectives, with the
latter corresponding to type IIB supergravity compactified on $S^5$.
When consistently truncated, this yields $\mathcal N=8$ gauged
supergravity in five dimensions.  Although this theory is generally
well understood, in many cases it is possible to further simplify the
system by invoking an $\mathcal N=2$ subsector of the full theory
which retains the $\mathcal N=2$ supergravity multiplet coupled to two
abelian vectors (often denoted the `STU model').

Three-charge black hole solutions in the STU model were first obtained
in \cite{Behrndt:1998ns,Behrndt:1998jd} as AdS generalizations of
asymptotically Minkowskian $R$-charged black holes.  However, in the
BPS limit these solutions in fact develop naked singularities; they
have been called superstars in \cite{Myers:2001aq} because of their
relation to distributions of giant gravitons.  Subsequently, genuine
BPS black holes were obtained by Gutowski and Reall in
\cite{Gutowski:2004ez,Gutowski:2004yv} by the addition of two equal
angular momenta, which were generalised to have arbitrary angular
momenta in \cite{cclp,Kunduri:2006ek}.  From a supergravity point of
view, this provides an explicit de-singularization of the superstar by
turning on rotation.

Following the work of Lin, Lunin and Maldacena on bubbling AdS
\cite{LLM}, it was shown that the 1/2 BPS superstar, corresponding to
specifying an intermediate boundary value on the AdS disk, may be
desingularized by an alternative distortion of the AdS disk into an
ellipse.  The resulting `AdS bubble' solution (including the
three-arbitrary charge generalization) was presented in
\cite{popebubble}, and involves additional $\mathcal N=8$ scalar
excitations which lie outside the conventional $\mathcal N=2$
truncation.  Furthermore, these AdS bubble solutions are horizon-free
and everywhere regular.  (The one, two and three charged bubbles
preserve $1/2$, $1/4$ and $1/8$ of the supersymmetries, and may be
described by ellipsoidal droplets in the generalized LLM phase space
\cite{recentliu}.)

Although these two methods for avoiding singularities are rather
distinct (one uses rotation to generate a horizon, while the other has
no horizon, but requires going beyond the STU model), they both apply
to the same system of IIB supergravity on AdS$_5\times S^5$.  Thus, in
this paper we wish to develop a unified framework for describing all
of the above BPS solutions in a five-dimensional supergravity context.
In order to do so, we have to add three additional scalars,
$\varphi_I$ with $I=1,2,3$, to the STU model.  These scalars arise
naturally from the diagonal elements of the $SL(6,\mathbb R)/SO(6)$
coset of the $S^5$ reduction of IIB supergravity to $\mathcal N=8$ in
five dimensions.  Viewed from a purely $\mathcal N=2$ perspective,
these scalars reside within three hypermultiplets.\footnote{However,
these are incomplete hypermultiplets, as we ignore their other
components.  This suffices for our present purposes, since we wish to
study supersymmetric configurations in which the other components of
the supermultiplets vanish.}  The unified picture we use is then that
of the STU model ($\mathcal N=2$ gauged supergravity with two vector
multiplets) coupled to three incomplete hypermultiplets.

The Gutowski-Reall black holes \cite{Gutowski:2004ez,Gutowski:2004yv}
were obtained using the $G$-structure (invariant tensor) method of
constructing supersymmetric solutions.  This method was initially
developed for minimal $\mathcal N=2$ supergravity in four dimensions
\cite{Tod:1983pm,Tod:1995jf}, and subsequently applied to minimal
ungauged \cite{Gauntlett:2002nw} and gauged \cite{Gauntlett:2003fk}
$\mathcal N=2$ supergravities in five dimensions.  One advantage of
the $G$-structure method is that it leads to a full classification (as
well as an implicit construction) of {\it all} backgrounds admitting
at least one Killing spinor.  In this way, one could in principle
obtain a complete understanding of all regular solutions of the STU
model coupled to hypermatter scalars $\varphi_I$, with or without
horizon.  In practice, however, the invariant tensor construction
which arises for this model is predicated on the choice of an
appropriate four-dimensional K\"ahler base upon which the rest of the
solution is built.  This choice of base leads to an extremely rich
structure of solutions, as can be witnessed from all the recent
developments in constructing new BPS black holes and black rings in
five dimensions.

In this paper, we limit ourselves to a cohomogeneity-one base with
bi-axial symmetry, which preserves $SU(2)_L\times U(1)\subset
SU(2)_L\times SU(2)_R\simeq SO(4)$ isometry.  This is sufficient to
obtain all known black holes and AdS bubbles with two equal rotations
turned on.  Curiously, however, the isometry of the base is not
required by the supersymmetry analysis to extend to that of the full
solution, a fact which was also noted in \cite{Figueras:2006xx} in the
context of cohomogeneity-two solutions.  As part of our analysis, we
find that solutions with the full $\mathbb R\times SU(2)_L\times U(1)$
isometry in five dimensions always admit a $U(1)$ breaking distortion,
leading to a distortion of AdS$_5$ at asymptotic infinity
\cite{behrndtklemm,gauntlett}.  Closed timelike curves (CTC's) may be
avoided in these G\"odel-like backgrounds, provided the distortion is
sufficiently small.

This paper is organized as follows. In section~2, we review gauged
${\cal N}=8$ supergravity in five dimensions, and discuss its
truncation to the STU model coupled to three incomplete
hypermultiplets. In section~3, we discuss the $G$-structure approach
to constructing supersymmetric backgrounds. In section~4, we present
the system of first-order equations for supersymmetric backgrounds
that preserve a time-like Killing vector and which have a bi-axial
four-dimensional K\"{a}hler base space. In section~5, we present some
explicit solutions which do not involve hyperscalars, such as black
holes, solitons and time machines. These solutions can be generalized
by relaxing the $SU(2)_L\times U(1)$ isometry to $SU(2)_L$. In
section~6, we discuss solutions which do involve hyperscalars, and
which are generalizations of the AdS bubbles \cite{popebubble}.  We
discuss bubbling generalizations of the Klemm-Sabra black holes in
section~7, and conclude in section~8.  Details regarding differential
identities for the invariant tensors, as well as the system of
equations governing a tri-axial four-dimensional K\"{a}hler base
space, are left for the appendices.

\section{Truncation of $\mathcal N=8$ supergravity}

Since we are interested in truncating $\mathcal N=8$ supergravity into
either matter coupled $\mathcal N=2$ supergravity or bosonic
subsectors thereof, we begin with the decomposition of the $\mathcal
N=8$ supergravity multiplet into $\mathcal N=2$ multiplets.  This is
presented in Table~\ref{tbl:decomp}, where we also give the lowest
weight energies $E_0$ and the representations under $SU(3)\times
U(1)\subset SU(4)$.  Here $U(1)$ is the $R$-symmetry of the $\mathcal
N=2$ theory embedded within the $SO(6)\simeq SU(4)$ $R$-symmetry of
the full $\mathcal N=8$ theory.

Note that the standard truncation of $\mathcal N=8$ to the STU model
($\mathcal N=2$ supergravity coupled to two vector multiplets)
corresponds to retaining two of the eight vectors in the maximal torus
of $SU(3)$.  In addition to gravity, the STU model has three abelian
vectors $A_\mu^I$ (one of which is the graviphoton) and two
unconstrained scalars, which may be traded off for three scalars $X^I$
satisfying the cubic constraint $X^1X^2X^3=1$.  Since this is a model
with vector multiplets, it is naturally described using very special
geometry.

In addition to the matter content of the STU model, we are interested
in retaining three additional scalars $\varphi_I$ of the $\mathcal
N=8$ theory.  {}From the $\mathcal N=8$ point of view, these
additional scalars share a common origin with the $X^I$ scalars as the
diagonal elements of the $SL(6,\mathbb R)/SO(6)$ coset representative
\begin{equation}
\mathcal M=\mbox{diag}(\sqrt{X^1}e^{\varphi_1/2},\sqrt{X^1}e^{-\varphi_1/2},
\sqrt{X^2}e^{\varphi_2/2},\sqrt{X^2}e^{-\varphi_2/2},\sqrt{X^3}
e^{\varphi_3/2},\sqrt{X^3}e^{-\varphi_3/2}),
\label{eq:xvpdiag}
\end{equation}
which is contained inside the $E_{6(6)}/USp(8)$ scalar manifold of
$\mathcal N=8$ supergravity.  However, despite this common origin, the
$\varphi_I$ scalars fall outside of the $\mathcal N=2$ vector
multiplets.  In particular, these additional scalars are parts of
hypermultiplets of the first type listed in Table~\ref{tbl:decomp}.
While, it is clear that they alone are insufficient to comprise the
bosonic parts of complete multiplets in themselves, the supersymmetry
analysis below nevertheless allows us to obtain solutions to the full
$\mathcal N=8$ theory in which only this restricted set of fields is
active.

\begin{table}[t]
\begin{center}
\begin{tabular}{llll}
$\mathcal N=2$ multiplet&fields&$E_0$ values&$SU(3)\times U(1)$\\
\hline
\noalign{\vspace*{4pt}}
graviton&$(h_{\mu\nu},\psi_\mu,A_\mu)$&$(4,\fft72,3)$&
$(\mathbf1_0,\mathbf 1_{\pm1},\mathbf 1_0)$\\[4pt]
gravitino&$(\psi_\mu,A_\mu,B_{\mu\nu},\lambda)$&
$(\fft72,3,3,\fft52)$&
$(\mathbf3_0,\mathbf 3_1,\mathbf3_{-1},\mathbf 3_0)_{1/3}$+conj.\\[4pt]
vector&$(A_\mu,\lambda,\phi)$&$(3,\fft52,2)$&
$(\mathbf8_0,\mathbf8_{\pm1},\mathbf8_0)$\\[4pt]
tensor&$(\lambda,B_{\mu\nu},\phi,\lambda)$&
$(\fft72,3,3,\fft52)$&
$(\mathbf3_0,\mathbf3_{-1},\mathbf3_{-1},\mathbf3_{-2})_{1/3}$+conj.\\[4pt]
hypermatter (1)&$(\phi,\lambda,\phi)$&$(3,\fft52,2)$&
$(\mathbf6_0,\mathbf6_{-1},\mathbf6_{-2})_{2/3}$+conj.\\[4pt]
hypermatter (2)&$(\phi,\lambda,\phi)$&$(4,\fft72,3)$&
$(\mathbf1_0,\mathbf1_{-1},\mathbf1_{-2})$+conj.
\end{tabular}
\end{center}
\caption{Decomposition of the $\mathcal N=8$ supergravity multiplet
into $\mathcal N=2$ multiplets under $SU(4)\supset SU(3)\times U(1)$.}
\label{tbl:decomp}
\end{table}

In principle, the addition of hypermatter requires us to consider the
full matter coupled $\mathcal N=2$ gauged supergravity
\cite{Ceresole:2000jd}.  However, for simplicity, we restrict
ourselves to the STU model coupled to the three additional $\varphi_I$
scalars.  As a result, we shall not need the entire machinery of
$\mathcal N=2$ matter couplings ({\it i.e.,} very special geometry for
vector multiplets and quaternionic geometry for hypermultiplets), but
will instead follow a direct reduction of the $\mathcal N=8$
expressions into their $\mathcal N=2$ counterparts.  We thus begin
with a review of the $\mathcal N=8$ theory, which serves as the
initial point of our analysis.

\subsection{The $\mathcal N=8$ supergravity}

Gauged $\mathcal N=8$ supergravity in five dimensions was constructed
in \cite{Pernici:1985ju,Gunaydin:1984qu,Gunaydin:1985cu}.  The bosonic
fields consist of the metric $g_{\mu\nu}$, $SO(8)$ adjoint gauge
fields $A_{\mu\,IJ}$, antisymmetric tensors $B_{\mu\nu}{}^{I\,\alpha}$
transforming as $(\mathbf6,\mathbf2)$ under $SO(6)\times SL(2,\mathbb
R)$ and 42 scalars $V_{AB}{}^{ab}$ parameterizing the coset
$E_{6(6)}/USp(8)$ and transforming as
$\mathbf{20}'+\mathbf{10}+\overline{\mathbf{10}} +\mathbf1+\mathbf1$
under $SO(6)$.  The fermions are the 8 gravitini $\psi_{\mu\,a}$ and
48 dilatini $\chi_{abc}$, all transforming under $USp(8)$.

Following the notation of \cite{Gunaydin:1984qu,Gunaydin:1985cu}, but
working in signature $(-,+,+,+,+)$, the gauged $\mathcal N=8$
Lagrangian has the form
\begin{equation}
e^{-1}\mathcal L=R-\ft16P_{\mu\,abcd}P^{\mu\,abcd}-\ft18H_{\mu\nu\,ab}
H^{\mu\nu\,ab}+\ft12\overline\psi_\mu{}^a\gamma^{\nu\nu\rho}D_\rho
\psi_{\rho\,a}+\ft1{12}\overline\chi^{abc}\gamma^\mu D_\mu\chi_{abc}
-V+\cdots,
\end{equation}
where we have only written the kinetic terms explicitly.  Here $D_\mu$
is the gravitational as well as $SL(6,\mathbb R)\times USp(8)$ covariant
derivative, and
\begin{equation}
P_\mu{}^{abcd}=\tilde V^{ab\,AB}D_\mu V_{AB}{}^{cd},\qquad
P_\mu{}^{abcd}\equiv P_\mu{}^{[abcd]|},
\label{eq:skin}
\end{equation}
is the scalar kinetic term.  The two-forms $H_{\mu\nu\,ab}$ are a
combination of the gauge fields and anti-symmetric tensors
\begin{equation}
H_{\mu\nu}{}^{ab}=F_{\mu\nu\,IJ}V^{IJ\,ab}
+B_{\mu\nu}{}^{I\alpha}V_{I\alpha}{}^{ab}.
\end{equation}
Finally, the scalar potential $V$ may be written in terms of the $W$-tensor
as
\begin{equation}
V=-\ft12g^2(2W_{ab}^2-W_{abcd}{}^2),\qquad
W_{abcd}=\epsilon^{\alpha\beta}\delta^{IJ}V_{I\alpha\,ab}V_{J\beta\,cd}.
\label{eq:spotwtens}
\end{equation}

To leading order, the supersymmetry transformations for the gravitini and
dilatini take the form
\begin{eqnarray}
\delta\psi_{\mu\,a}&=&D_\mu\epsilon_a+\ft{i}{12}(\gamma_\mu{}^{\nu\rho}
-4\delta_\mu^\nu\gamma^\rho)F_{\nu\rho\,ab}\epsilon^b-\ft{i}3g
\gamma_\mu W^c{}_{acb}\epsilon^b,\nonumber\\
\delta\chi_{abc}&=&-i\sqrt2\gamma^\mu P_{\mu\,abcd}\epsilon^d
+\ft3{4\sqrt2}\gamma^{\mu\nu}F_{\mu\nu\,[ab}\epsilon_{c]|}
+3\sqrt2gW_{d[abc]|}\epsilon^d.
\label{eq:susyvar}
\end{eqnarray}
Note that the $USp(8)$ indices $a,b,\ldots$ are raised and lowered
with the symplectic matrix $\Omega_{ab}$, and the symplectic-Majorana
Weyl spinors satisfy
\begin{equation}
\overline\lambda^a\equiv\lambda_a^\dagger\gamma^0=\Omega^{ab}\lambda_b^TC,
\end{equation}
where $C$ is the charge conjugation matrix.

Before considering the truncation to $\mathcal N=2$, however, we first
examine the scalar sector of the theory.  Although the complete
$\mathcal N=8$ scalar manifold is given by the coset
$E_{6(6)}/USp(8)$, the gauging of $SO(6)\subset USp(8)$ complicates
the explicit treatment of these scalars.  For this reason, we now
consider the simpler subsector of the scalar manifold corresponding to
taking $SL(6,\mathbb R)\times SL(2,\mathbb R)\subset E_{6(6)}$.
Furthermore, this subset of scalars has a natural Kaluza-Klein origin
from the $S^5$ reduction of IIB supergravity; the $\mathbf{20}'$
scalars living on $SL(6,\mathbb R)/SO(6)$ correspond to metric
deformations on $S^5$, while the $SL(2,\mathbb R)/SO(2)$ scalars
descend directly from the ten-dimensional IIB dilaton-axion.  In
particular, these $\mathbf{20}'$ scalars, along with the $SO(8)$ gauge
fields, were precisely the fields retained in the $S^5$ Pauli
reduction of \cite{Cvetic:2000nc}.  Note that, while this system is a
consistent bosonic truncation of $\mathcal N=8$ supergravity, it is
however not supersymmetric (even if fermions were to be included).
This is because the $\mathbf{20}'$ scalars, corresponding to $E_0=2$
in Table~\ref{tbl:decomp}, comprise only a subset of the first
hypermultiplet listed.  The remaining scalars in the hypermultiplet
originate from the reduction of the complexified three-form in IIB on
$S^5$.

Denoting the $SL(6,\mathbb R)/SO(6)$ and $SL(2,\mathbb R)/SO(2)$ coset
representatives by $\mathcal M^I{}_J$ and $\mathcal N^\alpha{}_\beta$,
respectively, we follow \cite{Gunaydin:1985cu} and obtain the $E_{6(6)}$
elements
\begin{equation}
U^{MN}{}_{IJ}=2\mathcal M^{-1\,[M}{}_{[I}\mathcal M^{-1\,N]}{}_{J]},\qquad
U_{J\beta}{}^{I\alpha}=\mathcal M^I{}_J\mathcal N^\alpha{}_\beta.
\label{eq:e66elem}
\end{equation}
Transforming to a $USp(8)$ basis using a set of imaginary antisymmetric
$SO(7)$ Dirac matrices $\Gamma_i$ ($i=0,1,\ldots,6$ while $I=1,\ldots,6$)
results in the coset representatives
\begin{equation}
V^{IJ\,ab}=\ft14(\Gamma_{KL})^{ab}\mathcal M^{-1\,I}{}_K\mathcal M^{-1\,J}_L,
\qquad
V_{I\alpha}{}^{ab}=\ft1{2\sqrt2}(\Gamma_{K\beta})^{ab}\mathcal M^K{}_I
\mathcal N^\beta{}_\alpha,
\end{equation}
along with the inverses
\begin{equation}
\tilde V_{IJ\,ab}=\ft12(\Gamma_{KL})^{ab}\mathcal M^K{}_J\mathcal M^L{}_J,
\qquad
\tilde V_{ab}{}^{I\alpha}=-\ft1{2\sqrt2}(\Gamma_{K\beta})^{ab}
\mathcal M^{-1\,I}{}_K\mathcal N^{-1\,\alpha}{}_\beta.
\end{equation}
In this case, the $W$-tensor of (\ref{eq:spotwtens}) reduces to
\begin{equation}
W_{abcd}=\ft{i}8[(\Gamma_I)_{ab}(\Gamma_J\Gamma_0)_{cd}-
(\Gamma_I\Gamma_0)_{ab} (\Gamma_J)_{cd}]M^{IJ},
\end{equation}
where
\begin{equation}
M^{IJ}=\mathcal M^I{}_K\mathcal M^J{}_L\delta^{KL},\quad\hbox{or}\quad
M=\mathcal M\mathcal M^T.
\label{eq:mcalm}
\end{equation}
Using
\begin{equation}
W_{ab}=-\ft14\delta_{ab}\Tr M,\qquad
(W_{abcd})^2=2\Tr(M^2),
\end{equation}
and substituting into (\ref{eq:spotwtens}) yields the scalar potential
\begin{equation}
V=-\ft12g^2[(\Tr M)^2-2\Tr(M^2)].
\label{eq:vpot}
\end{equation}
Note that the $SL(2,\mathbb R)$ scalars (or equivalently the IIB
dilaton-axion) do not enter the potential.

Continuing with this specialization of the scalar sector, we find that
the gauge fields enter in the combination
\begin{equation}
F_{\mu\nu}{}^{ab}=F_{\mu\nu\,IJ}V^{IJ\,ab}
=\ft14F_{\mu\nu\,IJ}(\Gamma_{KL})^{ab}\mathcal M^{-1\,I}{}_K
\mathcal M^{-1\,J}{}_L.
\end{equation}
The final quantity we need is the scalar kinetic term $P_\mu{}^{abcd}$
defined in (\ref{eq:skin}).  The condition that $P_\mu{}^{abcd}$ is
automatically symplectic-trace free determines the composite $USp(8)$
connection $Q_{\mu\,a}{}^b$ to be
\begin{equation}
Q_{\mu\,a}{}^b=\ft12(\Gamma_{IJ})^{ab}(\mathcal M\partial_\mu
\mathcal M^{-1})^I{}_J+\ft{i}2(\Gamma_0)^{ab}\epsilon_{\alpha\beta}
(\mathcal N\partial_\mu\mathcal N^{-1})^\alpha{}_\beta-\ft14gA_{\mu\,IJ}
(\Gamma_{KL})^{ab}\mathcal M^{-1\,I}{}_K\mathcal M^L{}_J.
\label{eq:usp8compc}
\end{equation}
This shows up both in the covariant derivative in the gravitino variation
and in the scalar kinetic term
\begin{eqnarray}
P_\mu{}^{abcd}&=&\tilde V^{ab}{}_{IJ}D_\mu V^{IJ\,cd}+2\tilde V^{ab\,I\alpha}
D_\mu V_{I\alpha}{}^{cd},\nonumber\\
&=&\tilde V^{ab}{}_{IJ}[\partial_\mu V^{IJ\,cd}-Q_{\mu\,e}{}^cV^{IJ\,ed}
-Q_{\mu\,e}{}^dV^{IJ\,ce}-2gA_{\mu\,IK}V^{KJ\,cd}]\nonumber\\
&&+2\tilde V^{ab\,I\alpha}[\partial_\mu V_{I\alpha}{}^{cd}-Q_{\mu\,e}{}^c
V_{I\alpha}{}^{ed}-Q_{\mu\,e}{}^dV_{I\alpha}{}^{ce}-gA_{\mu\,IJ}
V_{J\alpha}{}^{cd}]\nonumber\\
&=&\ft14(\Gamma_{I\beta})^{ab}(\Gamma_{I\alpha})^{cd}(\mathcal N\partial_\mu
\mathcal N^{-1})^\alpha{}_\beta\nonumber\\
&&+\ft14[(\Gamma_{IM})^{ab}(\Gamma_{JM})^{cd}+(\Gamma_J)^{ab}(\Gamma_I)^{cd}
-(\Gamma_J\Gamma_0)^{ab}(\Gamma_I\Gamma_0)^{cd}](\mathcal M\partial_\mu
\mathcal M^{-1})^I{}_J\nonumber\\
&&-\ft14gA_{\mu\,IJ}[(\Gamma_{KM})^{ab}(\Gamma_{LM})^{cd}
-(\Gamma_L)^{ab}(\Gamma_K)^{cd}+(\Gamma_L\Gamma_0)^{ab}(\Gamma_K\Gamma_0)^{cd}]
\mathcal M^K{}_I\mathcal M^{-1\,J}{}_L\nonumber\\
&&-[Q_\mu^{ac}\Omega^{bd}+Q_\mu^{bd}\Omega^{ac}
-Q_\mu^{ad}\Omega^{bc}-Q_\mu^{bc}\Omega^{ad}-\ft12Q_\mu^{cd}\Omega^{ab}].
\label{eq:skint}
\end{eqnarray}
These expressions, in principle, allow us to work out the full gravitino
and dilatino variations (\ref{eq:susyvar}) in terms of the explicit
parameterization (\ref{eq:e66elem}) of the $SL(6,\mathbb R)\times
SL(2,\mathbb R)$ scalars.

\subsection{The truncation to $\mathcal N=2$}
\label{sec:trunc}

As indicated in \cite{Gunaydin:1985cu}, the gauged $\mathcal N=8$
theory admits two maximal truncations to $\mathcal N=2$ supergravity.
The first retains only the hypermatter shown in Table~\ref{tbl:decomp}
coupled to the $\mathcal N=2$ graviton multiplet, while the second
corresponds to keeping only the vector and tensor multiplets.  A
further consistent truncation of this second case to the zero weight
sector of $SU(3)$ then yields the standard STU model, namely $\mathcal
N=2$ supergravity coupled to two vector multiplets.

We are mainly interested in a truncation of the above
$\mathcal N=8$ theory, where we retain the three gauge fields
\begin{equation}
A^1=A^{12},\qquad A^2=A^{34},\qquad A^3=A^{56},
\end{equation}
on the maximal torus of $SO(6)$, along with the five scalars
(\ref{eq:xvpdiag}) parameterizing the diagonal component of the
$SL(6;\mathbb R)/SO(6)$ coset.  Using (\ref{eq:mcalm}), we have
\begin{equation}
M=\mbox{diag}(X^1e^{\varphi_1},X^1e^{-\varphi_1},X^2e^{\varphi_2},
X^2e^{-\varphi_2},X^3e^{\varphi_3},X^3e^{-\varphi_3}),
\end{equation}
in which case $\Tr M=2\sum_I X^I\cosh\varphi_I$ and $\Tr
M^2=2\sum_I(X^I)^2 (\cosh^2\varphi_I+\sinh^2\varphi_I)$.  As a result,
from (\ref{eq:vpot}) we obtain the scalar potential
\begin{equation}
V=2g^2\left(\sum_I(X^I)^2\sinh^2\varphi_I
-2\sum_{I<J}X^IX^J\cosh\varphi_I\cosh\varphi_J\right).
\end{equation}
Note that this may be derived from a superpotential
\begin{equation}
W=g\sum_IX^I\cosh\varphi_I,
\label{eq:spot}
\end{equation}
using the relation
\begin{equation}
V=2\sum_\alpha(\partial_\alpha W)^2-\ft43W^2,
\end{equation}
where $\alpha=1,2,\ldots,5$ runs over the five unconstrained scalars.

After some manipulation of the scalar kinetic term (\ref{eq:skint}), we
find that the truncated bosonic action is
\begin{eqnarray}
e^{-1}\mathcal L&=&R-\ft12\partial\phi_\alpha^2-\ft12\partial\varphi_I^2
-\ft14(X^I)^{-2}(F_{\mu\nu}^I)^2-2g^2\sinh^2\varphi_I (A_\mu^I)^2
\nonumber\\
&&-V-\ft14\epsilon^{\mu\nu\rho\lambda\sigma}F_{\mu\nu}^1F_{\rho\lambda}^2
A_\sigma^3.
\label{eq:hyplag}
\end{eqnarray}
Note that the term proportional to $(A_\mu^I)^2$ originates from the
$SO(6)$ gauging in (\ref{eq:usp8compc}) and (\ref{eq:skint}).  The
lack of manifest gauge invariance in this action is a consequence of
the truncation to incomplete hypermultiplets.

In addition, the $\mathcal N=8$ supersymmetry transformations
decompose into four sets, each corresponding to a different embedding
of $\mathcal N=2$ into $\mathcal N=8$.  From a particular $\mathcal
N=2$ perspective, we may focus on a single set.  However, note that in
general the other three sets of supersymmetries may be completely
broken, unless additional symmetries are present beyond what is
imposed by the $\mathcal N=2$ analysis below.  For example,
three-charge non-rotating solutions preserve 1/2 of the $\mathcal N=2$
supersymmetries, but only 1/8 of the $\mathcal N=8$ ones
(corresponding to preserving four real supercharges in either case).

We end up with the $\mathcal N=2$ sector supersymmetry transformations
\begin{eqnarray}
\delta\psi_{\mu\,i}&=&\nabla_\mu\epsilon_i+
\fft{i}{24}(\gamma_\mu{}^{\nu\rho}
-4\delta_\mu^\nu\gamma^\rho)\mathcal F_{\nu\rho}\epsilon_i
+\ft12g\mathcal A_\mu\epsilon_{ij}\epsilon_j+
\ft{i}6W\gamma_\mu\epsilon_{ij}
\epsilon_j,\nonumber\\
\delta\lambda_{I\,i}&=&-i\gamma^\mu\partial_\mu\varphi_I\epsilon_i
+2ig\gamma^\mu A_\mu^I\sinh\varphi_I\epsilon_{ij}\epsilon_j
-2gX^I\sinh\varphi_I\epsilon_{ij}\epsilon_j,\nonumber\\
\delta\chi^{(1)}_i&=&-i\gamma^\mu\partial_\mu\log((X^1)^2/(X^2X^3))
\epsilon_i
-\ft12\gamma^{\mu\nu}(2(X^1)^{-1}F_{\mu\nu}^1-(X^2)^{-1}F_{\mu\nu}^2
-(X^3)^{-1}F_{\mu\nu}^3)\epsilon_i\nonumber\\
&&-2g(2X^1\cosh\varphi_1-X^2\cosh\varphi_2
-X^3\cosh\varphi_3)\epsilon_{ij}\epsilon_j,\nonumber\\
\delta\chi^{(2)}_i&=&-i\gamma^\mu\partial_\mu\log((X^2)^2/(X^1X^3))
\epsilon_i
-\ft12\gamma^{\mu\nu}(-(X^1)^{-1}F_{\mu\nu}^1+2(X^2)^{-1}F_{\mu\nu}^2
-(X^3)^{-1}F_{\mu\nu}^3)\epsilon_i\nonumber\\
&&-2g(-X^1\cosh\varphi_1+2X^2\cosh\varphi_2
-X^3\cosh\varphi_3)\epsilon_{ij}\epsilon_j,
\label{eq:susytrans}
\end{eqnarray}
where we have defined the graviphoton combinations
\begin{eqnarray}
\mathcal A_\mu&\equiv&A_\mu^1\cosh\varphi_1+A_\mu^2\cosh\varphi_2
+A_\mu^3\cosh\varphi_3,\nonumber\\
\mathcal F_{\mu\nu}&\equiv&(X^1)^{-1}F_{\mu\nu}^1+(X^2)^{-1}F_{\mu\nu}^2
+(X^3)^{-1}F_{\mu\nu}^3,
\label{eq:gphdef}
\end{eqnarray}
and where the superpotential $W$ is given in (\ref{eq:spot}).  The
spinors $\epsilon_i$, $i=1,2$ are now to be considered as $\mathcal
N=2$ spinors.

The gravitino and gaugino variations can almost be written in very
special geometry language (for the STU model) where, instead of taking
$V_I=1/3$, we use $V_I=\fft13\cosh\varphi_I$.  The $\varphi_I$ scalars
are parts of hypermultiplets and, when frozen to their constant values
$\varphi_I=0$, the gauging parameters $V_I$ take on their standard
constant values.

Note also that in the ungauged theory (obtained by taking $g\to0$),
the hypermultiplets decouple from the vector multiplets, at least in
the supersymmetry transformations.  This is just the standard
decoupling of $\mathcal N=2$ vector and hyper multiplets.
Furthermore, in the truncation to the dilatonic hypermultiplet scalars
$\varphi_I$, they also decouple from the gravitino multiplet.  (The
axionic ones will show up via the composite connection
$Q_{\mu\,a}{}^b$.)

\section{Supersymmetry analysis}
\label{sec:susyanal}

We shall use the invariant tensor approach for constructing
supersymmetric backgrounds.  This $G$-structure analysis has been
successfully applied to many systems, including minimal $\mathcal N=2$
supergravity in four dimensions \cite{Tod:1983pm,Tod:1995jf} as well
as minimal ungauged \cite{Gauntlett:2002nw} and gauged
\cite{Gauntlett:2003fk} $\mathcal N=2$ supergravities in five
dimensions.  The inclusion of vectors in the five-dimensional gauged
$\mathcal N=2$ case was investigated in
\cite{Gutowski:2004ez,Gutowski:2004yv} in the context of constructing
supersymmetric black holes.

We are of course interested in constructing supersymmetric backgrounds
where the hypermatter scalars $\varphi_I$ are active.  In this
context, the BPS conditions for obtaining static spherically symmetric
solutions were analyzed in \cite{Cacciatori:2002qx} for gauged
$\mathcal N=2$ supergravity coupled to hypermatter.  This was further
generalized in \cite{Celi:2003qk,Cacciatori:2004qm} for the complete
system including both vector and hypermultiplets.  (See also
\cite{Bellorin} for a complete analysis of ungauged supergravity
coupled to hypermatter.)  These studies, however, assumed spherical
symmetry from the outset, an assumption that we wish to relax.  Thus,
we shall mainly follow the invariant tensor procedure of
\cite{Gauntlett:2002nw,Gauntlett:2003fk,Gutowski:2004ez,Gutowski:2004yv}.
This procedure starts with a construction of all tensors formed as
bilinears of the Killing spinor $\epsilon_i$ followed by an
examination of algebraic and differential identities related to these
tensors, which we now consider.

\subsection{Spinor bilinear identities}

Note that $\epsilon_i$ is an $\mathcal N=2$ symplectic-Majorana
spinor, with $i$ an $Sp(2)\simeq SU(2)$ index.  In particular, it
carries eight real spinor components.  We may form a complete set of
real bilinears
\begin{equation}
f=\ft{i}2\overline\epsilon^i\epsilon_i,\qquad
K_\mu=\ft12\overline\epsilon^i\gamma_\mu\epsilon_i,\qquad
\Phi_{\mu\nu}^a=\ft12\overline\epsilon^i(\tau^a)_{ij}\gamma_{\mu}\epsilon_j,
\label{eq:bilins}
\end{equation}
where $\tau^a$ are the usual Pauli matrices.  We take as a convention
$\fft1{5!}\epsilon_{\mu\nu\rho\lambda\sigma}
\gamma^{\mu\nu\rho\lambda\sigma}=i$
along with $\epsilon_{01234}=1$.

The standard Fierz identities give the normalization relations
\begin{equation}
K^2=-f^2,\qquad (\Phi_{\mu\nu}^a)^2=12f^2,
\label{eq:ksqfierz}
\end{equation}
along with
\begin{eqnarray}
i_K\Phi^a&=&0,\nonumber\\
i_K*\Phi^a&=&-f\Phi^a,\nonumber\\
\Phi^a\wedge \Phi^b&=&-2\delta^{ab}f*K,\nonumber\\
\Phi^a_{\lambda\mu}\Phi^{b\,\lambda}{}_\nu&=&\delta^{ab}(f^2g_{\mu\nu}
+K_\mu K_\nu)-\epsilon^{abc}f\Phi_{\mu\nu}^c,
\label{eq:morefierz}
\end{eqnarray}
where for any $p$-form $\omega$ we define $(i_K\omega)_{\mu_1\cdots
\mu_{p-1}} =K^\nu \omega_{\nu \mu_1\cdots \mu_{p-1}}$.  These
identities indicate that the set $(K,\Phi^2)$ defines a preferred
$U(2)$ structure.  In the ungauged case, the addition of $\Phi^1$ and
$\Phi^3$ would yield a preferred $SU(2)$ structure.  However, here
they are charged under the gauged $U(1)$, and hence are only covariant
and not invariant.

Integrability of the $U(2)$ structure may be investigated through the
differential identities which arise from the supersymmetry variations.
These are presented in Appendix~\ref{sec:diffid}.  As usual,
symmetrization of the $\nabla_\mu K_\nu$ identity arising from the
gravitino variation (\ref{eq:gravids}) demonstrates that $K^\mu$ is a
Killing vector:
\begin{equation}
2\nabla_{(\mu}K_{\nu)}=0.
\label{eq:kkilling}
\end{equation}
This, combined with (\ref{eq:ksqfierz}), ensures that $K^\mu$ is an
everywhere non-spacelike Killing vector.  Since we are interested in
constructing black holes (and related solitonic bubbles), we take the
timelike case where
\begin{equation}
K^2=-f^2<0.
\end{equation}

\subsection{Specializing the metric}

We now assume $K^\mu$ is a timelike Killing vector with norm
$K^2=-f^2$ where $f\ne0$.  For simplicity of notation, we take $f>0$.
(The $f<0$ case is similar, and involves a modified choice of signs.
However, it does not give rise to any intrinsically new solutions.)
In this case, we may specialize the metric to be of the form
\begin{equation}
ds^2=-f^2(dt+\omega)^2+f^{-1}h_{mn}dx^m\,dx^n.
\label{eq:metspec}
\end{equation}
Note that we take $K=\partial/\partial t$, so that
$K=-f^2(dt+\omega)=-fe^0$ where $e^0=f(dt+\omega)$.

Given that $i_K\Phi^a=0$ from (\ref{eq:morefierz}), we see that the
two-forms $\Phi^a$ live on the four-dimensional base with metric
$h_{mn}$.  The remaining identities in (\ref{eq:morefierz}) are then
equivalent to
\begin{equation}
*_4\Phi^a=-\Phi^a,\qquad
\Phi^a\wedge\Phi^b=-2\delta^{ab}*_41,\qquad
\Phi^a_{mp}\Phi^b_{nq}h^{pq}=\delta^{ab}h_{mn}-\epsilon^{abc}\Phi^c_{mn}.
\end{equation}
This indicates that the three $\Phi^a$ form a set of anti-self-dual
2-forms on the base that satisfy the algebra of unit quaternions.  In
the ungauged case, this is sufficient to demonstrate a preferred
$SU(2)$ structure; here $\Phi^2$ defines a $U(2)$ structure, while
$\Phi^1$ and $\Phi^3$ are charged under the gauged $U(1)$.

To make the structure explicit, we define the canonical 2-form $J$
along with a complex 2-form $\Omega$ according to
\begin{equation}
J=\Phi^2,\qquad\Omega=\Phi^1+i\Phi^3.
\end{equation}
This set $(J,\Omega)$ determines the $U(2)$ structure on the base,
\begin{equation}
J\wedge\Omega=0,\qquad
J\wedge J=\ft12\Omega\wedge\Omega^*=-2*_41.
\end{equation}
Integrability of $J$ and $\Omega$ will be taken up below, when we
consider the differential identities.

\subsection{Determining the gauge fields}

In order to obtain a supersymmetric background, we need to determine
not only the metric $g_{\mu\nu}$ (or equivalently the quantities $f$,
$\omega$ and $h_{mn}$) but also the matter fields $A^I$, $X^I$ and
$\varphi_I$.  We begin with the gauge fields.  Firstly, using
(\ref{eq:aicond}), which we take as either a gauge condition (when
$g\sinh\varphi_I=0$) or as a consequence of the hyperino
transformations, we may write the potentials as
\begin{equation}
A^I=f^{-1}X^IK+\beta^I=-X^Ie^0+\beta^I,
\label{eq:potdecomp}
\end{equation}
where $\beta^I$ lives exclusively on the base ({\it i.e.,} $i_K\beta^I=0$).
The field strengths are then
\begin{equation}
F^I=dA^I=-d(X^Ie^0)+d\beta^I=f^{-2}d(fX^I)\wedge K-fX^Id\omega+d\beta^I.
\label{eq:fi=da}
\end{equation}
Note that the one-form identity (\ref{eq:1formid}) is automatically
satisfied.

To proceed, we may turn to the two-form identities (\ref{eq:2formid}).
For simplicity, we define the components $\overline F^I$ of the field
strengths on the base by writing (\ref{eq:fi=da}) as
\begin{equation}
F^I=f^{-2}d(fX^I)\wedge K+\overline F^I.
\end{equation}
The two-form identities then reduce to
\begin{equation}
*_4\left((X^I)^{-1}\overline F^I\right)+(X^J)^{-1}\overline F^J
+(X^K)^{-1}\overline F^K=-fd\omega+2gf^{-1}JX^I\cosh\varphi_I,
\end{equation}
where $I\ne J\ne K$.  By breaking this up into self-dual and anti-self
dual parts, we obtain a complete determination of the anti-self dual
components
\begin{equation}
\left((X^I)^{-1}\overline F^I\right)^-=-f(d\omega)^-+gf^{-1}J
(X^J\cosh\varphi_J+X^K\cosh\varphi_K)
\label{eq:fasdc}
\end{equation}
and a single condition on the sum of the self-dual components
\begin{equation}
\left((X^1)^{-1}\overline F^1\right)^+
+\left((X^2)^{-1}\overline F^2\right)^+
+\left((X^3)^{-1}\overline F^3\right)^+=-f(d\omega)^+.
\end{equation}
In terms of $d\beta^I$, these conditions become
\begin{equation}
(d\beta^I)^-=gf^{-1}J\left(\fft1{X^J}\cosh\varphi_K+\fft1{X^K}\cosh\varphi_J
\right),
\label{eq:fasdcond}
\end{equation}
and
\begin{equation}
\fft1{X^1}(d\beta^1)^+
+\fft1{X^2}(d\beta^2)^+
+\fft1{X^3}(d\beta^3)^+=2f(d\omega)^+.
\label{eq:fsdcond}
\end{equation}

To show that the base metric $h_{mn}$ is K\"ahler, we note from the
first equation of (\ref{eq:3formid}) that $dJ=0$ is trivially
satisfied.  In order to examine $d\Omega$, we decompose the
graviphoton $\mathcal A$ defined in (\ref{eq:gphdef}) into timelike
and spatial components using (\ref{eq:potdecomp}).  If we multiply
(\ref{eq:aicond}) by $\cosh\varphi_I$ and sum over $I$, we see that
the graviphoton necessarily satisfies the condition
\begin{equation}
gi_K\mathcal A=-fW.
\end{equation}
This ensures that the timelike component of $\mathcal A$ cancels against
the superpotential term in (\ref{eq:3formid}), leaving
\begin{equation}
d\Omega=-ig(\beta^1\cosh\varphi_1+\beta^2\cosh\varphi_2+\beta^3\cosh\varphi_3)
\wedge\Omega.
\end{equation}
Combined with $dJ=0$, we see that the base is indeed
K\"ahler\footnote{The conditions for K\"ahlerity can be expressed as
$dJ=0$, $J\wedge\Omega=0$, $d\Omega=\im \omega\wedge\Omega$ for some
1-form $\omega$.  The 1-form $\omega$ is arbitrary up to the addition
of any $(0,1)$-form.  There exists a choice for $\omega$ such that
$d\omega ={\cal R}$, the Ricci form.}, with Ricci form satisfying
\begin{equation}
\mathcal R=-g\, d(\beta^1\cosh\varphi_1+\beta^2\cosh\varphi_2
+\beta^3\cosh\varphi_3).
\end{equation}

It is now easy to see that the remaining 0-form gaugino identities in
(\ref{eq:0formid}) are satisfied.  Furthermore, with some work, we may
also verify that the additional 3-form identities (\ref{eq:3formid})
are satisfied as well.  Note, in particular, that the identities
related to $J$ ({\it i.e.,} the $a=2$ identities) require that the
graviphoton-free combinations of $\overline F^I$ be $(1,1)$-forms on
the base
\begin{equation}
J_{[m}{}^p\overline F_{n]p}^{(\alpha)}=0.
\end{equation}
This is trivially satisfied because the self-dual part of $\overline
F^{(\alpha)}$ is automatically $(1,1)$, while from (\ref{eq:fasdc}) we
see that the graviphoton-free anti-self-dual part is proportional to
$J$, which is itself a $(1,1)$-form.  We have not explicitly checked
the 4-form identities (\ref{eq:4formid}), but expect them to hold
without any new conditions.

\subsection{Determining the hypermatter scalars}
\label{sec:hypers}

So far, other than using (\ref{eq:aicond}) to determine the time
component of $A^I$, we have not focused on the hypermatter scalars
$\varphi_I$.  Thus, the above analysis is essentially identical to
that of \cite{Gauntlett:2003fk,Gutowski:2004ez,Gutowski:2004yv} for
minimal gauged $\mathcal N=2$ supergravity and gauged $\mathcal N=2$
supergravity coupled to vector multiplets.  However, we now turn to
the hyperino identities (\ref{eq:hyperid}).  The zero-form identities
have already been accounted for, so we proceed directly with the
1-form identity, which requires that $\varphi_I$ live on the
four-dimensional base, and satisfy
\begin{equation}
d\varphi_I=-2g\sinh\varphi_IJ_m{}^n\beta_n^Idx^m.
\label{eq:dvarphi}
\end{equation}
This relates the hypermultiplet scalars $\varphi_I$ with the spatial
components of the gauge fields $\beta^I$.  Note that this can
equivalently be written as
\begin{equation}
J\wedge d\varphi_I=2g\sinh\varphi_I*_4\beta^I.
\label{eq:hypercond}
\end{equation}

As it turns out, this condition is sufficient to ensure that all the
remaining hyperino identities are satisfied.  To see this, we may turn
directly to the supersymmetry transformation $\delta\lambda_{I\,i}$
given in (\ref{eq:susytrans}).  Substituting in (\ref{eq:dvarphi}) as
well as the gauge field decomposition (\ref{eq:potdecomp}) gives
\begin{equation}
\delta\lambda_{I\,i}=2ig\sinh\varphi_I\beta_n\gamma^m\epsilon_{ij}
[\delta_m^n\delta_j^k-J_m{}^n\epsilon_{jk}]\epsilon_k
-2gX^I\sinh\varphi_I\epsilon_{ij}[1+i\gamma^0]\epsilon_j.
\label{eq:delhypsol}
\end{equation}
This expression must vanish in order for $\epsilon_i$ to be a Killing
spinor.  So long as $g\sinh\varphi_I\ne0$, the second term in
(\ref{eq:delhypsol}) yields the familiar condition
\begin{equation}
i\gamma^0\epsilon_i=-\epsilon_i.
\label{eq:12bpsproj}
\end{equation}
If this were the only condition, then the solution would be 1/2 BPS.
However, we must also ensure the vanishing of the first term in
(\ref{eq:delhypsol}).  This may be accomplished by noting that, so
long as $\beta_n$ is generic, we must demand
\begin{equation}
\gamma^m[\delta_m^n\delta_i^j-J_m{}^n\epsilon_{ij}]\epsilon_j=0.
\end{equation}
Multiplying on the left by $\fft14\gamma_n$ then gives
\begin{equation}
[\delta_i^j+\ft14(J\cdot\gamma)\epsilon_{ij}]\epsilon_j=0.
\label{eq:14bpsproj}
\end{equation}
Since $(J\cdot\gamma)\epsilon_{ij}$ has eigenvalues $\pm4,0,0$,
we see that this yields a 1/4 BPS projection.  Furthermore, since
$J$ is anti-self dual:
\begin{equation}
[(J\cdot\gamma)i\tau^2]^2=8(1+\gamma^{1234})=8(1-i\gamma^0)
\end{equation}
we see that the projection (\ref{eq:14bpsproj}) is compatible with
(\ref{eq:12bpsproj}), and hence the complete system remains 1/4 BPS
when both projections inherent in (\ref{eq:delhypsol}) are taken into
account.

\subsection{Completing the solution}

To complete the solution, we must impose the $F^I$ equations of
motion.  Note that by making the ansatz (\ref{eq:potdecomp}) on the
gauge potential, we are guaranteed to satisfy the Bianchi identities.
From (\ref{eq:hyplag}), the $F^I$ equation of motion reads
\begin{equation}
d\left(*\fft1{(X^I)^2}F^I\right)=F^J\wedge F^K-4g^2\sinh^2\varphi_I*A^I.
\label{eq:feom}
\end{equation}
Using the explicit forms for $A^I$ and $F^I$ given in
(\ref{eq:potdecomp}) and (\ref{eq:fi=da}), we see that this equation
decomposes into one whose component lies along $e^0$, and one which
only resides on the base.  The former turns out to be trivially
satisfied, provided the supersymmetry conditions (\ref{eq:fasdcond}),
(\ref{eq:fsdcond}) and (\ref{eq:hypercond}) hold.  On the other hand,
the part of (\ref{eq:feom}) which lies on the base gives rise to the
second-order equation
\begin{equation}
d*_4d\left(\fft1{fX^I}\right)=-d\beta^J\wedge d\beta^K
+2g\cosh\varphi_I d\omega\wedge J+4g^2\sinh^2\varphi_If^{-2}X^I*_41.
\end{equation}
This suggests that we introduce three independent functions
\begin{equation}
H_I=\fft1{fX^I},
\end{equation}
so that the second-order equation of motion becomes
\begin{equation}
d*_4dH_I=-d\beta^J\wedge d\beta^K
+2g\cosh\varphi_I d\omega\wedge J+4g^2\sinh^2\varphi_IH_JH_K*_41.
\label{eq:newheom}
\end{equation}
Note that the constraint $X^1X^2X^3=1$ indicates that the function $f$
is given by
\begin{equation}
f=(H_1H_2H_3)^{-1/3}.
\end{equation}

We have now found all of the constraints arising from supersymmetry
and the equations of motion. To summarize, the solution is given by
the metric
\begin{equation}
ds^2=-(H_1H_2H_3)^{-2/3}(dt+\omega)^2+(H_1H_2H_3)^{1/3}h_{mn}dx^m dx^n,
\end{equation}
gauge potentials
\begin{equation}
A^I=-\fft1{H_I}(dt+\omega)+\beta^I,
\label{eq:gaugepot}
\end{equation}
vector multiplet scalars
\begin{equation}
X^I=\fft{(H_1H_2H_3)^{1/3}}{H_I},
\end{equation}
and hypermultiplet scalars $\varphi_I$.  The metric $h_{mn}$ on the
base is K\"ahler, with anti-self-dual K\"ahler form $J$ and
holomorphic $(2,0)$-form $\Omega$.  The remaining quantities
$(\varphi_I,\omega,\beta^I)$ must satisfy
\begin{eqnarray}
(d\beta^I)^-&=&gJ(H_J\cosh\varphi_K+H_K\cosh\varphi_J),\nonumber\\
2d\omega^+&=&H_1(d\beta^1)^++H_2(d\beta^2)^++H_3(d\beta^3)^+,\nonumber\\
\mathcal R&=& -g\, d(\beta^1\cosh\varphi_1+\beta^2\cosh\varphi_2
+\beta^3\cosh\varphi_3),\nonumber\\
d\varphi_I&=&-2g\sinh\varphi_IJ_m{}^n\beta_n^Idx^m,
\label{eq:finsusycond}
\end{eqnarray}
as well as the equations of motion (\ref{eq:newheom}), which we repeat here:
\begin{equation}
d*_4dH_I=-d\beta^J\wedge d\beta^K
+2g\cosh\varphi_I d\omega\wedge J+4g^2\sinh^2\varphi_IH_JH_K*_41.
\label{eq:fineom}
\end{equation}
%

\section{Supersymmetric solutions}
\label{sec:susysols}

{}From the above analysis, we see that the starting point for
constructing supersymmetric solutions is the choice for the
four-dimensional K\"ahler base. In this paper, we shall focus on the
bi-axial case.  However, for completeness, the first-order equations
for the most general tri-axial ansatz for a cohomogeneity-one solution
with $S^3$ orbits are presented in Appendix~\ref{sec:triaxial}. In the
bi-axial case, a gauge can be chosen such that the K\"ahler metric on
the base is cast into the form
\begin{equation}
ds_4^2=\fft{dx^2}{4xh(x)}+\fft{x}4(\sigma_1^2+\sigma_2^2+h(x)\sigma_3^2),
\label{eq:basemet}
\end{equation}
where $\sigma_i$ are $SU(2)$ left-invariant 1-forms satisfying
$d\sigma_1=-\sigma_2\wedge\sigma_3$.  Corresponding to this metric, we
introduce a natural vierbein basis
\begin{equation}
e^1=\fft{dx}{2\sqrt{xh}},\qquad e^2=\fft{\sqrt x}2\sigma_1,\qquad
e^3=\fft{\sqrt x}2\sigma_2,\qquad e^4=\fft{\sqrt{xh}}2\sigma_3.
\label{eq:vierbein}
\end{equation}
This base admits an anti-self-dual K\"ahler form
\begin{equation}
J=\ft14d(x\sigma_3)=e^1\wedge e^4-e^2\wedge e^3,
\label{eq:bhkahf}
\end{equation}
and has the Ricci form
\begin{equation}
\mathcal R=d\Big((2-xh' -2h)\sigma_3\Big)=2\Big( h'+\fft2x(h-1)\Big)\, e^2\wedge e^3-
2(xh''+3h')\, e^1\wedge e^4.
\end{equation}

In addition to the K\"ahler metric on the base, we also make an ansatz
for the 1-form $\omega$, as well as the gauge functions $\beta^I$,
\begin{equation}
\omega = w_1\sigma_1+w_2\sigma_2+w_3\sigma_3,\qquad
\beta^I=U_1^I\sigma_1+U_2^I\sigma_2+U_3^I\sigma_3.
\label{eq:omegbetI}
\end{equation}
A true bi-axial solution, such as the black holes of
\cite{Gutowski:2004ez,Gutowski:2004yv}, will have only the components
proportional to $\sigma_3$ turned on.  However, by allowing
non-trivial $\sigma_1$ and $\sigma_2$ components, we may also develop
solutions asymptotic to deformed AdS$_5$, as investigated in
\cite{behrndtklemm,gauntlett}.  Note that while the base metric
(\ref{eq:basemet}) preserves $SU(2)_L\times U(1)$ isometry, the
complete five-dimensional solution only preserves a reduced $SU(2)_L$
isometry unless all the $\sigma_1$ and $\sigma_2$ components vanish in
(\ref{eq:omegbetI}).

We find that $d\omega$ decomposes into self-dual and anti-self-dual
components according to
\bea
(d\omega)^{\pm} &=&
2\sqrt{h}\Big(w_1'\mp\fft{w_1}{xh}\Big)(e^1\wedge e^2\pm e^3\wedge e^4)
+2\sqrt{h}\Big(w_2'\mp\fft{w_2}{xh}\Big)(e^1\wedge e^3\mp e^2\wedge e^4)\nn\\
&&+2\Big(w_3'\mp\fft{w_3}{x}\Big)(e^1\wedge e^4\pm e^2\wedge e^3).
\eea
Similarly, $(d\beta^I)^\pm$ has the same form as $(d\omega)^{\pm}$,
except with $w_i\rightarrow U_i^I$.  In this case, the first-order
supersymmetry equations (\ref{eq:finsusycond}) (or equivalently the
first-order tri-axial equations (\ref{firstordereqns1})) reduce to
\begin{eqnarray}
\varphi_I'&=&-\fft{2g}{xh}U_3^I\sinh\varphi_I,\nonumber\\
(xU_3^I)'&=&\fft{gx}2(H_J\cosh\varphi_K+H_K\cosh\varphi_J),\nonumber\\
U_j^{I'} &=& -\fft{U_j^I}{xh},\nn\\
\left(\fft{w_3}x\right)'&=&\ft12\sum_IH_I\left(\fft{U_3^I}x\right)',
\nonumber\\
w_j^{'}-\fft{w_j}{xh} &=&\sum_I H_I U_j^{I'},\nn\\
(x^2h)'&=&2x+2gx\sum_IU_3^I\cosh\varphi_I,
\label{firstordereqns}
\end{eqnarray}
as well as the algebraic conditions
\be
\sum_I U_j^I \cosh\varphi_I=0\,,\qquad gU_j^I\sinh\varphi_I=0,
\label{eq:algcons}
\ee
where $j=1,2$.  The second-order equation of motion
(\ref{eq:fineom}) can be expressed as
\begin{equation}
0=\Bigl[x^2hH_I'+4\sum_{i=1}^3 U_i^JU_i^K\Bigr]'
-2g\cosh\varphi_I(xw_3)'+g^2\sinh^2\varphi_IxH_JH_K.
\end{equation}
This may be rewritten as
\begin{equation}
0=\Bigl[x^2hH_I'+4\sum_{i=1}^3 U_i^JU_i^K-2g\cosh\varphi_Ixw_3\Bigr]'
+g^2\sinh^2\varphi_I\left(xH_JH_K-4\fft{w_3}hU_3^I\right),
\label{secondordereqn}
\end{equation}
where we have used the first-order equation for $\varphi_I$.

Note that, just as in \cite{Gutowski:2004ez,Gutowski:2004yv}, we
could have chosen the opposite sign for the K\"ahler form in
(\ref{eq:bhkahf}).  This simply corresponds to taking
\begin{equation}
w_i\to-w_i,\qquad U_i^I\to-U_i^I,
\end{equation}
in the expressions above.

\section{Solutions without hyperscalars}

We are principally interested in obtaining and classifying all
solutions of the supersymmetric bi-axial system given by the
first-order equations (\ref{firstordereqns}), algebraic constraints
(\ref{eq:algcons}) and equation of motion (\ref{secondordereqn}).  To
proceed, we first consider the case when the hypermatter scalars
$\varphi_I$ are set to zero.  This case corresponds to the gauged
supergravity version of the STU model, and has been extensively
studied.  Nevertheless, as shown below, there are still surprises to
be found when analyzing these solutions.

By setting $\varphi_I=0$, the above system of equations reduces to
\begin{eqnarray}
(xU_3^I)'&=&\fft{gx}2(H_J+H_K),\nonumber\\
\left(\fft{w_3}x\right)'&=&\ft12\sum_IH_I\left(\fft{U_3^I}x\right)',
\nonumber\\
(x^2h)'&=&2x+2gx\sum_IU_3^I,
\label{eq:novphi1st}
\end{eqnarray}
involving the $\sigma_3$ components, and
\begin{eqnarray}
U_j^{I'}&=&-\fft{U_j^I}{xf},\nonumber\\
w_j'-\fft{w_j}{xh}&=&\sum_IH_IU_j^{I'},\nonumber\\
0&=&\sum_IU_j^I
\label{eq:novphi2nd}
\end{eqnarray}
($j=1,2$) involving the $\sigma_1$ and $\sigma_2$ components.  In
addition, the second-order equation reduces to
\begin{equation}
0=\Bigl[x^2hH_I'+4\sum_{i=1}^3U_i^JU_i^K-2gxw_3\Bigr]',
\label{eq:novphieom}
\end{equation}
which admits a first integral that is proportional to the Noether
electric charge $Q^I$ of the gauge fields.  Note that this equation of
motion is the only expression coupling the $\sigma_1$ and $\sigma_2$
components $U_j^I$ to the functions $H_I$.

We may generate a formal solution to the above system by assuming the
functions $H_I$ to be arbitrary.  The functions $U_3^I$, $w_3$ and $h$
can then be obtained by successive integration of the first-order
equations in (\ref{eq:novphi1st}).  Similarly, the functions $U_j^I$
and $w_j$ follow from (\ref{eq:novphi2nd}) by integration.  At this
stage, all quantities may now be formally written in terms of $H_I$
and its integrals.  Inserting these expressions into the
(\ref{eq:novphieom}) then gives rise to a set of integro-differential
equations whose solutions correspond to generically 1/4 BPS
configurations solving all equations of motion.  However, in practice,
such a formal solution is difficult to analyze.  Hence, we instead
turn to some explicit solutions.

\subsection{Solutions with $\mathbb R\times SU(2)_L\times U(1)$ isometry}

We recall that the bi-axial ansatz (\ref{eq:basemet}) involves a
K\"ahler base with $SU(2)_L\times U(1)$ isometry.  This isometry may
be extended to the complete solution by taking $U^I_1=0=U^I_2$ and
$w_1=0=w_2$, in which case the equations (\ref{eq:novphi2nd}) are
trivially satisfied.  Together with time translational invariance, the
full isometry of the solution is $\mathbb R\times SU(2)_L\times U(1)$.

     Even in this case, however, an analytic form for the general
solution is not apparent.  Nevertheless, by assuming `harmonic
functions' of the form $H_I=1+q_I/x$, we find a class of solutions
given by
\bea
H_I &=& 1 + \fft{q_I}{x},\nn\\
U^I_3 &=& \ft12 g\, (x + q_J + q_K) + \fft{\alpha_I}{x},\nn\\
h&=&1 + g^2 (x + \sum_I q_I) +\fft{2g\sum_I \alpha_I}x +
\fft{\gamma}{x^2},\nn\\
w_3&=&\ft12 g (x + \sum_I q_I)+\fft{2\sum_I \alpha_I+
g \sum_{I<J}q_Iq_J}{4x}
+\fft{\sum_I q_I\, \alpha_I}{3x^2}.
\label{soln1}
\eea
This solution is parameterized by the quantities $q_I$ and $\alpha_I$,
$I=1,2,3$ satisfying the condition
\begin{equation}
q_1\,\alpha_1=q_2\,\alpha_2=q_3\,\alpha_3.\label{alphacon}
\end{equation}
In this case, the constant $\gamma$ may be expressed as
\be
\gamma=\fft{4\alpha_I\alpha_J}{q_K},
\ee
for any choice of $I\ne J\ne K$, so long as $q_K$ is non-vanishing.
(If all three charges $q_I$ vanish, then $\gamma$ is arbitrary.)

Alternatively, this solution can be reexpressed in terms of the $H_I$
functions as
\begin{eqnarray}
H_I &=& 1 + \fft{q_I}{x},\nn\\
U^I_3 &=& \fft{g}{2}x H_J H_K+\fft{\gamma_I}{x},\nn\\
h&=& 1+g^2x\prod_IH_I+\fft{2g\sum_I\gamma_I}{x}
+\fft{4(g\gamma_1q_1+\gamma_2\gamma_3/q_1)}{x^2},\nn\\
w_3&=& \fft{g}{2}x \prod_I H_I+\fft{\sum_I \gamma_I}{2x}
+\fft{\gamma_1 q_1}{x^2},
\label{soln1a}
\end{eqnarray}
where
\begin{equation}
\gamma_I\equiv\alpha_I-\ft12gq_Jq_K.
\end{equation}
Note that the integration constants satisfy
\be
\qquad q_1\,\gamma_1=q_2\,\gamma_2=q_3\,\gamma_3.
\ee
As a result, the last terms in the expressions for $h$ and $w_3$ are in fact
symmetric in the charges.  As we shall see, both of the above sets
of expressions will be useful for exploring various limits as well
as generalizations of the solutions.

These solutions generically preserve $1/4$ of the supersymmetry
of the $D=5$, ${\cal N}=2$ gauged supergravity.
The mass, angular momentum and $R$-charges are given by
\bea
M &=&2gJ + \ft14 (q_1 + q_2 + q_3) -
\ft14 g (\alpha_1 + \alpha_2 + \alpha_3) +
\ft18 g^2 (q_1q_2 + q_1 q_3 + q_2q_3),\nn\\
J &=& -\ft14 (\alpha_1 + \alpha_2 + \alpha_3) + \ft18 g(2 \gamma + q_1q_2 +
q_1q_3 + q_2 q_3) - \ft13 g^2 (\alpha_1 q_1 + \alpha_2 q_2
+ \alpha_3q_3)\nn\\
&&+\ft14 g^3 q_1 q_2 q_3,\nn\\
Q_I&=& \ft14 q_I - \ft14 g (\alpha_J + \alpha_K-\alpha_I)
+ \ft18g^2(q_I (q_J + q_K) - q_J q_K),\label{mass1}
\eea
or equivalently
\begin{eqnarray}
M&=&2gJ+\ft14(q_1+q_2+q_3)-\ft14 g(\gamma_1+\gamma_2+\gamma_3),\nonumber\\
J&=&-\ft14(\gamma_1+\gamma_2+\gamma_3)+g\gamma_2\gamma_3/q_1,\nonumber\\
Q_I&=&\ft14q_I-\ft14g(\gamma_J+\gamma_K-\gamma_I).\label{mass2}
\end{eqnarray}
Note that in presenting the mass, charge and angular momentum results,
we suppress a common factor that is the volume of spatial
principal orbits,
which can be $S^3$, or a lens space $S^3/{\mathbb Z}_k$, for some
integer $k$, which is fixed by a specific regularity requirement of the
solutions.  It is easy to see that these quantities satisfy the
BPS condition
\be
M=2gJ + Q_1 + Q_2 + Q_3\,.
\label{BPScondition}
\ee
It should be noted that black holes in five dimensions may carry two
independent angular momenta, $J_1$ and $J_2$.  Our choice of a
cohomogeneity-one base, however, restricts the system to two equal 
angular momenta, $J_1=J_2=J$.  In general, the solution becomes non-rotating 
when $\alpha_I = \ft12 g q_J q_K$, or equivalently when $\gamma_I=0$.

The non-rotating solutions with $\gamma_I=0$ are in fact the original
superstars of \cite{Behrndt:1998ns,Behrndt:1998jd}.  These have naked
singularities at $x=0$.  On the other hand, the supersymmetric black
holes of Gutowski and Reall \cite{Gutowski:2004ez,Gutowski:2004yv} are
recovered when $\alpha_I=0$.  In this case, the radial coordinate $x$
runs from the horizon at $x=0$, where the geometry is a direct product
of AdS$_2$ and a squashed $S^3$, to asymptotic AdS$_5$ as
$x\to\infty$.  The three-equal-charge case of the solution
(\ref{soln1}) was found in \cite{behrndtklemm}, while the general case
was obtained in \cite{timemachines}.

In general, the solution (\ref{soln1}) describes a spacetime in which 
there is a region with closed timelike curves (CTC's).  Such a 
spacetime is sometimes referred to as a `time machine.' In this
case, $x$ runs from $x_0>0$, where $x_0$ is the greatest root of $f$, to
asymptotic infinity.  These time-machine solutions can be made perfectly
regular with appropriate assignments of the periodicity for the real time
coordinate $t$, as discussed in \cite{timemachines}.  Naked CTC's can be
avoided by imposing the additional condition that $w_3(x_0)=0$. 
This leads to the
supersymmetric solitons that are discussed below.

\subsubsection{Massless solitons}

    The properties of the solitons are largely determined by the
parameters $q_I$.  We shall first consider the case of the solution
given by (\ref{soln1}) with only a single $U(1)$ gauge field active.
This corresponds to having $q_2=0=q_3$, $\alpha_1=0$ and
$\alpha_2=\alpha_3\equiv c_1$.  Let us choose the parameters $q_1$
and $c_1$ so that
\be
h(x_0)=0\,,\qquad w_3(x_0)=0\,.\label{conditions}
\ee
The first condition is needed in order to avoid power-law curvature
singularities, while the second one ensures that there are no CTC's,
as we have discussed earlier.  These conditions can be satisfied by
setting
\be
q_1 = - \fft{g^2 x_0^2}{1 + g^2 x_0}\,,\qquad
c_1 = -\fft{g\,x_0^2}{2(1 + g^2 x_0)}\,.\label{masslesscon}
\ee
This implies that $q_1=2g\, c_1$.  Now we have
\bea
H_1&=&\fft{x + g^2 x_0 (x-x_0)}{x+g^2 x\, x_0}\,,\nn\\
w_3&=&\fft{g(x-x_0)(x+x_0 + g^2 x\, x_0)}{2x(1+g^2 x_0)}\,,\nn\\
h&=&\fft{(x-x_0)(x+x_0 + g^2 x\, x_0)(1+g^2 x)}{x^2(1 + g^2 x_0)}\,,
\eea
and indeed $x_0>0$ is the greatest root of $f$.  It follows that
the solution does not have a power-law curvature singularity for
$x\ge x_0$ with $x_0>0$.  In addition, there are no CTC's since we have
\be
g_{\psi \psi} = 
\fft{(x-x_0)(x + x_0 + g^2 x\, x_0)}{4x (1 + g^2 x_0)H_1^{2/3}}\ge 0\,.
\ee
The consequence of this is that $t$ is a globally defined time
coordinate, in that for any constant $t$, the spacetime is foliated by 
spatial sections.

      In order for the ($x$,$\psi$) subspace to form a smooth
$\R^2$ at $x=x_0$, the period of the angular coordinate $\psi$ must be
\be
\Delta\psi=\fft{4\pi}{2 + g^2 x_0}.
\label{period1}
\ee
In addition, in order for the level surfaces of the principal orbits
to be regular, the period of $\psi$ must be such that
\be
\Delta\psi=\fft{4\pi}{k}\,,
\ee
for some integer $k$.
As a consequence, the principal orbits are lens spaces $S^3/{\mathbb Z}_k$.
Therefore, in order to avoid a conical singularity, $x_0$ is fixed to be
\be
x_0=\fft{k-2}{g^2}\,,
\ee
for each lens space $S^3/{\mathbb Z}_k$.  The requirement of $x_0>0$
implies that we must have $k\ge 3$. 

      It is easy to verify using (\ref{mass1}) that the mass, charge
and angular momentum all vanish for this soliton, when the conditions
(\ref{masslesscon}) for the regularity and the absence of CTC's are
imposed.  In this sense, it provides an explicit example of a
`texture' in gauged supergravity. Let us be more precise about this,
since from the gravitational point of view one can always add an
arbitrary constant to the mass. Throughout this paper, we shall take
the mass $M_{\rm AdS}$ of the AdS vacuum to be zero, since the CFT
Casimir energy is not relevant for our discussion. Then, by zero mass
we mean specifically that $M=M_{\rm AdS}$.

\subsubsection{Massive solitons}

We now consider the case for which $q_3=0$ and $q_1$ and $q_2$ are
nonvanishing. Then we must have $\alpha_1=\alpha_2=0$ by virtue of
(\ref{alphacon}). Let us choose the parameters $q_1$, $q_2$ and
$\alpha_3$ such that the conditions given by (\ref{conditions}) are
satisfied. This can be achieved by setting
\be
g^2=\fft{x_0}{(x_0+q_1)(x_0 +q_2)}\,,\qquad
\alpha_3 =-\ft12g(2x_0^2 + 2x_0(q_1 + q_2) + q_1 q_2)\,.\label{galpha3}
\ee
It follows that we have
\bea
H_1&=& 1+\fft{q_1}{x}\,,\qquad
H_2= 1 + \fft{q_2}{x}\,,\qquad H_3=1\,,\nn\\
h &=& \fft{(x-x_0)[x x_0 + 2x_0^2 + 2x_0(q_1 + q_2) +
q_1 q_2]}{x(x_0+q_1)(x_0+q_2)}\,,\nn\\
w_3 &=& \fft{(x-x_0)(x+x_0 + q_1 + q_2)}{2x}\,.
\eea
We can verify that the solution does not have a power-law curvature
singularity for $x\ge x_0$, where $x_0>\max\{0,-q_1,-q_2\}$.
There are no CTC's either, since we have
\be
g_{\psi \psi} = \fft{(x-x_0)[x^2 + x(x_0+q_1 + q_2) +
(x_0+q_1)(x_0+q_2)]}{4x^2 (H_1H_2)^{2/3}} \ge 0\,.
\ee
The ($x$,$\psi$) subspace forms an $\R^2$ near $x=x_0$ if the period
of $\psi$ is
\be
\Delta\psi=\fft{4\pi(x_0+q_1)(x_0+q_2)}{
3x_0^2 + 2 x_0(q_1+ q_2) + q_1 q_2}\,.\label{period2}
\ee
In order for the level surfaces of the principal orbits to be regular,
{\it i.e.,} $S^3/{\mathbb Z}_k$,
the period of the angle $\psi$ has to be $\Delta\psi=\fft{4\pi}{k}$.
Thus, we have
\be
k=\fft{3x_0^2 + 2 x_0(q_1+ q_2) + q_1 q_2}{(x_0+q_1)(x_0+q_2)}\,.
\ee
Note that there is no solution for $k=\pm 1$ that satisfies the
regularity conditions. For $k=2$, we have $x_0=\sqrt{q_1q_2}$ which,
together with (\ref{galpha3}), implies that $q_1$ and $q_2$ must both
be positive.  The other values of $k$ can only be achieved with at
least one of the $q_i$'s negative.

    The mass, charge and angular momentum for this solitonic solution
are given by
\bea
J &=& \fft{x_0}{4g}\,,\qquad M=\ft14 (3x_0+q_1 + q_2)\,.\nn\\
Q_1 &=& \ft14(q_1-x_0)\,,\qquad Q_2=\ft14(q_2-x_0)\,,\qquad
Q_3=\fft{x_0}{4}\,,
\eea
which of course satisfy the BPS condition (\ref{BPScondition}). This
charged rotating soliton has a positive mass.

\subsubsection{Negative mass solitons}

Finally, we consider the case in which none of the $q_i$ vanish. We
can take $\alpha_I=g\beta/q_I$, for a constant $\beta$. In order for
the conditions given by (\ref{conditions}) to be satisfied, we take
\bea
g &=& \fft{x_0\prod_{I<J}q_Iq_J+2q_1q_2q_3}{\sqrt{\prod_I
(x_0+q_I)(\prod_{I<J}q_I^2q_J^2-2q_1q_2q_3(2x_0+\sum_I q_I))}}\,,\nn\\
\beta &=& -\fft{q_1q_2q_3x_0(2x_0^2+2x_0 \sum_I q_I+
\prod_{I<J} q_Iq_J)}{2(x_0\prod_{I<J}q_Iq_J+2q_1q_2q_3)}\,.
\eea
The local expressions for this class of solutions were obtained in
\cite{timemachines} by taking the BPS limit of the non-extremal
rotating black hole solutions constructed in \cite{clp1,clp2}.  Here,
we analyse the solutions in more detail, and demonstrate that smooth
solutions with negative mass can also arise.  Since the resulting
expressions for the metric functions are rather long, we shall examine
only a couple of particular cases.

\bigskip
\noindent\underline{Single charge}
\bigskip

We first consider $q_2=q_3=2q_1\equiv -2q$, in which case the expressions
become significant simpler.  The soliton condition
(\ref{conditions}) implies that
\be
x_0=2q + \fft{2\sqrt{q}}{g}\,,\qquad \beta=\fft{2q^2}{g^2}(g^2 q-1)\,,
\label{schcon}
\ee
which requires that $q>0$. Consequently, we have
\bea
H_1 &=& 1-\fft{q}{x}\,,\qquad H_2=H_3=1-\fft{2q}{x}\,,\nn\\
h &=& \fft{(x+x_0-4q)(4q(x-x_0)+x_0^2)(x-x_0)}{(x_0-2q)^2 x^2}\,,\nn\\
w &=& -\fft{\sqrt{q} (x-q)(x+x_0-4q)(x-x_0)}{(x_0-2q)x^2}\,.
\eea
Thus, the solutions do not have a power-law curvature singularity
for $x\ge x_0$ and  $0<q<x_0/2$.  There are also no CTC's, since we have
\be
g_{\psi\psi}=\fft{(x-q)(x+x_0-4q)(x-x_0)}{4x^2(H_1H_2H_3)^{2/3}}\ge 0\,.
\ee
The ($x$,$\psi$) subspace forms an $\R^2$ near $x=x_0$ if the period
of $\psi$ is
\be
\Delta\psi=\fft{2\pi(x_0-2q)}{x_0}\,.
\ee
Combining this with the usual requirement that $\Delta\psi=4\pi/k$ and
with the condition (\ref{schcon}) yields
\be
x_0=\fft{k(k-2)}{2g^2}\,,\qquad q=\fft{(k-2)^2}{4g^2}\,.
\ee
Thus, we must have $k\ge 3$.

For these solitonic solutions, $M=Q_1=-q/4$ and $Q_2=Q_3=J=0$.  Thus,
we see that although all three $U(1)$ gauge fields $A_\1^I$ are turned
on, there is only one charge.  The solution has zero angular momentum
although it has rotations.  It is furthermore rather surprising that
regularity and the absence of CTC's implies that these solitons have
negative mass, or more specifically $M<M_{\rm AdS}$.  Solutions with
negative mass have been referred to as `phantom matter,' whose
repulsive behavior may be useful for modeling the observed
acceleration of the scale factor $a(t)$ of the universe
\cite{phantom1,phantom2}.

Positive mass theorems in general relativity have established that
asymptotically AdS solutions of the Einstein equations with physically
acceptable matter sources cannot have negative total mass.  For
instance, the negative mass Schwarzschild solution has a naked
power-law curvature singularity. Our solutions evade such positive
mass theorems by having an asymptotic geometry of AdS$_5/\Z_k$ with
$k\ge 3$, rather than AdS$_5$. To be more precise, the $S^3$ within
AdS$_5$ has been replaced by the lens space $S^3/\Z_k$.  Since these
solutions are supersymmetric, they are perturbatively stable against
local energy fluctuations.

\bigskip
\noindent\underline{Three equal charges}
\bigskip

Another simple example is the case of three equal charges, for which
we can set $q_I\equiv q$. While this case has already been discussed
in detail in \cite{timemachines}, the possibility of negative mass
solitons was not realized.  Again, after imposing the condition
(\ref{conditions}) we have
\be
g=\fft{\sqrt{-q}(2q+3x_0)}{(q+x_0)\sqrt{(q+x_0)(3q+4x_0)}},\qquad
\beta=\fft{qx_0(3q^2+6qx_0+2x_0^2)}{4q+6x_0},
\ee
and
\be
H_1 = H_2=H_3=1+\fft{q}{x}\,,\nn
\ee
\be
h = \fft{[-q(3x_0+2q)^2x^2+(3q^2+6qx_0+2x_0^2)(2x_0^2-3qx_0-3q^2)x+
(3q^2+6qx_0+2x_0^2)^2x_0] (x-x_0)}{(4x_0+3q)(x_0+q)^3x^2},\nn
\ee
\be
w = \fft{\sqrt{-q}[q(2x^2+11x_0x+2x_0^2+3q^2)+3(2q^2+x_0x)(x+x_0)]
(x-x_0)}{2(q+x_0)\sqrt{(q+x_0)(3q+4x_0)}\, x^2}\,.\qquad\qquad\qquad\qquad
\ee
There are two cases for which the solution is real and completely regular
for $x\ge x_0$: 
\bea
{\rm Case\ I}: && x_0<0\,,\qquad q>0\,,\qquad -q<x_0<-\ft34q\,,\nn\\
{\rm Case\ II}: && x_0>0\,,\qquad q<0\,,\qquad -q<x_0\,.
\label{twocases}
\eea
Note that the possibility that $x_0$ can be negative for regular
solutions can only arise when all three charge parameters $q_I$ are
non-vanishing, since otherwise negative $x$ would lead to a power-law
curvature singularity at $x=0$.  This can be seen easily by noting
that the radius square of the $S^2$ of the base space is given by
$x(H_1H_2H_3)^{1/3}$, which is a constant at $x=0$ for the case with
$q_I$ all non-vanishing, whilst becomes zero for the cases when at
least one of the $q_I$ vanishes.

      For both cases (\ref{twocases}), we have verified that there
are no CTC's since
\be
g_{\psi\psi}=\fft{[(4x_0+3q)x^2+(4x_0+3q)(x_0+3q)x+
(9qx_0+5q^2+3x_0^2)q](x-x_0)}{4(4x_0+3q)(x+q)^2}\ge 0\,.
\ee
The ($x$,$\psi$) subspace forms an $\R^2$ near $x=x_0$ if the period
of $\psi$ is
\be
\Delta\psi= \fft{4\pi (x_0+q)(4x_0+3q)}{(8x_0+5q)x_0} \,.
\ee
Combining this with the requirement that $\Delta\psi= 4\pi/k$ yields
\begin{equation}
x_0 =\begin{cases}
x_0^{\pm}= \fft{5-7k\pm\sqrt{(k+1)(k+25)}}{8(k-2)}q,&\mbox{for } k\ne 2;\\
-\ft23 q,&\mbox{for } k=2.\end{cases}
\end{equation}
In order to satisfy the conditions for Case I, we find that $k\le
-25$.  Interestingly enough, these conditions are met for either sign
in $x_0$.  The boundaries $x_0=-q$ and $x_0=-\ft34 q$ are saturated
for $k\rightarrow -\infty$ by $x_0=x_0^-$ and $x_0=x_0^+$,
respectively. On the other hand, the conditions for Case II are
satisfied only for $x_0=x_0^-$ with $k\ge 3$. Then the boundary
$x_0=-q$ is saturated for $k\rightarrow +\infty$.

For these solitonic solutions, 
\be
M = \fft{(12x_0^2+5q^2+15qx_0)q}{4(4x_0+3q)^2}\,,\qquad
J = -\ft14 \left(-\fft{(x_0+q)q}{4x_0+3q}\right)^{3/2}\,,\qquad
Q_i = \fft{(x_0+q)q}{4(4x_0+3q)}\,.
\ee
These three-equal charge solitons have positive mass for Case I and
negative mass for Case II.  As in the previous case, the negative-mass
solitons evade the positive mass theorems by being asymptotically
AdS$_5/{\mathbb Z}_k$, where $k\ge 3$.

\subsection{Solutions with $\mathbb R\times SU(2)_L$ isometry}

Returning to the first-order equations (\ref{eq:novphi1st}) and
(\ref{eq:novphi2nd}), we see that the above system always admits an
$\mathbb R\times SU(2)_L\times U(1)$ breaking deformation where $w_1$
and $w_2$ (multiplying $\sigma_1$ and $\sigma_2$ in the time
fibration, respectively) are turned on.  By keeping $U_1^I=0=U_2^I$,
this deformation is essentially restricted to the metric.  In
particular, the equation of motion (\ref{eq:novphieom}) is left
unchanged.  This deformation reduces the $\mathbb R\times
SU(2)_L\times U(1)$ isometry of the five-dimensional metric to
$\mathbb R\times SU(2)_L$ only, although the K\"ahler base is
undeformed and retains the full original isometry.

Integrating the second equation in (\ref{eq:novphi2nd}), we see that
the solution given by (\ref{soln1}) (or equivalently (\ref{soln1a})) can
be further generalized to include $\sigma_1$ and $\sigma_2$ in the
timelike fibration as follows:
\be
w_1 = c_1 u\,,\qquad w_2 = c_2 u\,,
\qquad u=u_0\,\exp\Big[ \int_{x_0}^x \fft{dx'}{x'f(x')}\Big]\,.
\ee
These solutions still preserve $1/4$ of the supersymmetry of the
$D=5$, ${\cal N}=2$ gauged supergravity.  For the $w_i$ generalisation
of the AdS rotating black holes, corresponding to $q_I$ all equal and
$\alpha_I=0$, this reduces to a family of solutions constructed in
\cite{behrndtklemm}. Furthermore, for $c_2=0$ and $q_I=\alpha_I=0$,
these solutions reduce to the deformations of AdS$_5$ constructed in
\cite{Gauntlett:2003fk} and further analyzed in \cite{gauntlett}.  We
can extend those solutions to include $c_2$, for which the absence of
CTC's in the G\"odel-like universe at asymptotic infinity can be
achieved by requiring
\be
c_1^2+c_2^2\le \fft{1}{4g^2}\,.\label{c1c2cond}
\ee
In obtaining this result, we have normalized $u$ by choosing an
appropriate $u_0$ such that $u=1$ for $x=\infty$.  We shall use the
same normalization for $u$ for other solutions as well. The
criteria for avoiding CTC's at $x=x_0$ are the same as for the
solutions in the previous subsection.

For general $q_I$, $u$ can be expressed in terms of a sum of
polynomial roots.  Note that, for solutions where $x$ runs from $x=0$
to $\infty$, the function $u$ runs from 0 at $x=0$ to 1 at $x=\infty$.
It is clear that there are no CTC's near $x=0$, while CTC's at large
$x$ can also be avoided by taking the condition (\ref{c1c2cond}).  For
example, for the $w_i$ generalisation of rotating black holes, we have
\be
u=\Big( 1 + \fft{q_1 + q_2 + q_3 + g^{-2}}{x}\Big) ^{
-\ft{1}{1 + g^2 (q_1 + q_2 + q_3)}}
\,.
\ee
It is easy to verify that there are no CTC's provided that
(\ref{c1c2cond}) is satisfied.  For solutions where $x$ runs from
$x=x_0>0$ to $\infty$, the function $u$ runs from 0 at $x=x_0$ to 1 at
$x=\infty$. Thus, again there are no CTC's near $x=x_0$ and the
condition for the absence of CTC's at infinity is the same as
(\ref{c1c2cond}).  The general expression for $u$ can be complicated.
In the special case of vanishing $q_3$, we find a simple expression,
given by
\be
u=\Big( \fft{A-B-2g^2 x}{A+B+2g^2 x}\Big)^{1/A}\,,
\ee
where
\be
A \equiv \sqrt{1+2g^2(q_1+q_2)+g^4(q_1-q_2)^2-8g^3 \gamma_3}
\,,\qquad B \equiv 1+g^2(q_1+q_2)\,.
\ee
Finally, we present the explicit expression for $u$ for the massless
soliton studied in section 5.1.1.  It is given by
\be
u=\fft{g^2(x-x_0)^\ft1{2+g^2 x_0}(x + \fft{x_0}{1 + g^2 x_0})^
\ft{1+g^2 x_0}{2 + g^2 x_0}}{1 + g^2 x}\,.
\ee

\section{Solutions with hyperscalars}

In this section, we study supersymmetric solutions with the
hypermatter scalars $\varphi_I$ turned on.  We obtain some new
explicit analytical solutions as well as a class of new numerical
solutions.

\subsection{Solutions with $\mathbb R\times SU(2)_L$ isometry}

      In \cite{popebubble}, a general class of static bubble solutions
were obtained for the STU model coupled to the three hypermatter
scalars $\varphi_I$.  Even with these additional scalars turned on,
the first-order equations (\ref{firstordereqns}) still allow $w_1$ and
$w_2$ to be turned on without affecting the equation of motion
(\ref{secondordereqn}).  As a result, this class of bubble solutions
admits a $SU(2)_L\times U(1)$ breaking deformation of the form
\bea
\cosh\varphi_I &=& (x H_I)',\nn\\
h &=& 1+g^2 x H_1H_2H_3,\nn\\
U_3^I &=& \fft{g}{2}x H_J H_K,\qquad U_1^I = U_2^I=0,\nn\\
w_3 &=& \fft{g}{2}x H_1H_2H_3,\qquad
w_1 = c_1 u,\qquad w_2 = c_2 u,
\eea
where the functions $H_I$ satisfy the equations
\be
h(x H_I)^{''}=-g^2 [(x H_I)^{'2}-1](H_1H_2H_3)H_I^{-1}\,,\label{Heqn}
\ee
and where the metric deformation function is given by
\be
u=u_0\,\exp\Big[ \int_{x_0}^x \fft{dx'}{x'h(x')}\Big].
\ee
For $c_1=c_2=0$, this reduces to the AdS bubbles constructed in
\cite{popebubble}, which generalize a subset of 1/2 BPS LLM solutions
\cite{LLM} to 1/4 and 1/8 BPS solutions by turning on two and three
independent $U(1)$ fields, respectively.  By relaxing the
$SU(2)_L\times U(1)$ isometry of the AdS bubbles \cite{popebubble} to
$SU(2)_L$ only, we find that there is a more general family of
solutions, for which the $c_1$ and $c_2$ deformation parameters are
non-zero.

   In the single-charge case, with $H_2=H_3=1$,
there is an explicit expression for $H_1$:
\be
H_1=\sqrt{1+\fft{2(1+g^2 q_1)}{g^2 x}+\fft{c^2}{g^4 x^2}}-
\fft{1}{g^2x}\,.
\label{singleH}
\ee
Regularity of the AdS bubble requires that $c=1$. In this case, the
deformation function $u$ becomes
\be
u=\fft{g^2(2+g^2q_1) x}{1+g^2(1+g^2q_1)x+\sqrt{(1+g^2x)^2+2g^4q_1 x}}\,.
\ee
The geometry runs from a timelike bundle over $\R^4$ at short distance
($x=0$) to a G\"odel-like universe asymptotically
($x\rightarrow\infty$).  Since $u$ and $w_3$ vanish linearly at $x=0$,
it follows that the solution does not have CTC's near $x=0$.  When
$x\rightarrow\infty$, the absence of CTC's requires the same
condition as in (\ref{c1c2cond}).
It is straightforward to verify that there are no CTC's from $x=0$ to
$\infty$ when the above condition is satisfied.

     For the generic three-charge situation, the equations
(\ref{Heqn}) do not seem to allow solutions to be found explicitly.
(The numerical analysis was performed in \cite{nonsusybubbles}.)
However, it is easy to see that the structure of the three-charge
solution is rather similar to that of the single-charge case.  The
coordinate runs from $x=0$ to $\infty$, with $H_I$ and $f$ being
certain constants at $x=0$.  It follows that the $w_i$ vanish at
$x=0$, implying no CTC's near $x=0$.  Since $H_I\sim 1 + q_I/x$ for
large $x$, for sufficiently small $c_1^2 + c_2^2$, CTC's can be
avoided in the asymptotic region. Then such bubbling solitonic
solutions are completely regular and are free of CTC's.

\subsection{General rotating bubbles with $\mathbb R\times SU(2)_L
\times U(1)$ isometry}

In the previous subsection, we obtained analytical solutions by imposing
the condition $\cosh\varphi_I=(x H_I)'$, which was originally given in
\cite{popebubble}.  Although this is a necessary condition for static
bubbles, it is not a direct consequence of supersymmetry and hence can
be relaxed when the system is rotating.  Here we consider the general
system with non-vanishing hypermatter scalars $\varphi_I$.  However,
we restrict our attention to solutions with $\mathbb R\times SU(2)_L
\times U(1)$ isometry such that the metric can be expressed in a
non-rotating frame in the asymptotic region.  This corresponds to
setting $U^I_1=U^I_2=0$ and $w_1=w_2=0$, but leaving all other fields
free up to the first-order equations (\ref{firstordereqns})
\begin{eqnarray}
\varphi_I'&=&-\fft{2g}{xh}U_3^I\sinh\varphi_I,\nonumber\\
(xU_3^I)'&=&\fft{gx}2(H_J\cosh\varphi_K+H_K\cosh\varphi_J),\nonumber\\
\left(\fft{w_3}x\right)'&=&\ft12\sum_IH_I\left(\fft{U_3^I}x\right)',\nn\\
(x^2h)'&=&2x+2gx\sum_IU_3^I\cosh\varphi_I,
\label{eq:geneq1}
\end{eqnarray}
and second-order equations (\ref{secondordereqn})
\begin{equation}
0=\Bigl[x^2hH_I'+4U_3^JU_3^K-2g\cosh\varphi_Ixw_3\Bigr]'
+g^2\sinh^2\varphi_I\left(xH_JH_K-4\fft{w_3}hU_3^I\right).
\label{eq:geneq2}
\end{equation}

         In general, there are two types of solitonic solutions.  The
first type can be referred to as $\R^4$ solitons, where the coordinate
$x$ runs from 0 to $\infty$.  This is because the geometry near $x=0$
is a direct product of time and $\R^4$.  It can be demonstrated
numerically that such solutions exist.  In order to do so, we may
first show that the above system admits a regular Taylor series
solution near $x=0$ of the form
\bea
H_I&=&h_I^0-\fft{g^2x}4\Bigl[(h_I^0)^2\cosh\varphi_J^0\cosh\varphi_K^0
+h_J^0h_K^0(1+3\sinh^2\varphi_I^0)\nn\\
&&\kern4.5em+\cosh\varphi_I^0(h_I^0(h_J^0\cosh\varphi_K^0
+h_K^0\cosh\varphi_J^0)-8\gamma)\Bigr]+\cdots,\nn\\
\cosh\varphi_I&=&\cosh\varphi_I^0-\fft{g^2x}2\sinh^2\varphi_I^0
(h_J^0\cosh\varphi_K^0+h_K^0\cosh\varphi_J^0)+\cdots,\nn\\
U_3^I&=&\fft{gx}4\Bigl[h_J^0\cosh\varphi_K^0+h_K^0\cosh\varphi_J^0\Bigr]
-\fft{g^2x^2}{24}\Bigl[-16\gamma\cosh\varphi_J^0\cosh\varphi_K^0\nn\\
&&\kern4em+2h_I^0(h_J^0\cosh\varphi_J^0(1+3\sinh^2\varphi_K^0)
+h_K^0\cosh\varphi_K^0(1+3\sinh^2\varphi_J^0))\nn\\
&&\kern4em+\cosh\varphi_I^0\bigl(
(h_J^0)^2(1+3\sinh^2\varphi_K^0)+(h_K^0)^2(1+3\sinh^2\varphi_J^0)\nn\\
&&\kern9em+2h_J^0h_K^0\cosh\varphi_J^0\cosh\varphi_K^0\bigr)\Bigr]+\cdots,
\eea
along with
\bea
h&=&1+\fft{g^2x}3\sum_Ih_I^0\cosh\varphi_J^0\cosh\varphi_K^0+\cdots,\nn\\
w_3&=&g\gamma x-\fft{g^3x^2}{48}\Bigl[-16\gamma\sum_I h_I^0
\cosh\varphi_J^0\cosh\varphi_K^0
+3\prod_I(h_J^0\cosh\varphi_K^0+h_K^0\cosh\varphi_J^0)\nn\\
&&\kern6em+6\sum_Ih_I^0\cosh\varphi_I^0((h_J^0)^2\sinh^2\varphi_K^0
+(h_K^0)^2\sinh^2\varphi_J^0)\Bigr]+\cdots.
\eea
In general, the solution to the system (\ref{eq:geneq1}) and
(\ref{eq:geneq2}) may be specified by 14 independent parameters.
However, regularity at the origin reduces this to the 7 parameters
$(\varphi_I^0, h_I^0, \gamma)$.  Numerical integration may then be
used to connect this solution to its most general counterpart
developed around $x=\infty$.

Since we find that the asymptotic solution at $x=\infty$ is well
behaved, we see that regular bubbling solutions may be obtained for
generic values of the 7 parameters $(\varphi_I^0, h_I^0, \gamma)$.
However, it should be noted that logarithmic terms are almost always
present in the expansion.  The presence of such terms gives rise to
potentially infinite mass for these rotating solitons.  Since mass
can be extracted from the behavior of the metric at the asymptotic
boundary, this infinite mass is closely related to
deformations of the $S^3$ at infinity which in turn leads to a
deformation of the global spacetime away from asymptotic AdS$_5$.

Furthermore, while smooth bubbling solutions exist for a large range
of parameters, they generally contain CTC's.  However, by adjusting
the initial parameters $(\varphi_I^0, h_I^0, \gamma)$ appropriately,
we find that solutions without CTC's may be obtained.

We now present the analysis for the single charge $\mathbb R^4$
soliton in somewhat more detail.  To obtain a single charge solution,
we set $H_2=H_3=1$ as well as $\varphi_2=\varphi_3=0$.  To avoid
generating magnetic field components for $A^2$ and $A^3$, we must also
set $U^2_3=U^3_3=w_3$.  The remaining non-trivial fields may then be
given in terms of two functions $H(x)$ and $\zeta(x)$:
\begin{eqnarray}
H_1&=&H,\nn\\
\cosh\varphi_1&=&(xH)'+\fft1x(x^2\zeta)'',\nn\\
U_3^1&=&\fft{gx}2,\nn\\
w_3&=&\fft{gx}2H+\fft{g}{2x}(x^2\zeta)',\nn\\
h&=&1+g^2xH+\fft{g^2}{x^2}(x^3\zeta)'.
\end{eqnarray}
In this single charge case, the combination of the first and second
order equations above reduce to a coupled set of two equations which
contain up to second derivatives of $H$ and third derivatives of
$\zeta$.  This indicates that the general solution may be specified by
five parameters.  An expansion at infinity gives
\begin{eqnarray}
H&=&1+\fft1{g^2x}(h_1+h_{11}\log x)+\fft1{g^4x^2}h_2+\cdots,\nn\\
g^2\zeta&=&-\ft12h_{11}
+\fft1{g^2x}(f_1+h_{11}(1-\ft12h_{11})\log x)
+\fft1{g^4x^2}(f_2+h_{11}(1-\ft12h_{11})^2\log x)+\cdots,\qquad
\label{eq:asyexp}
\end{eqnarray}
where the asymptotic parameters are $(h_1,h_{11},h_2,f_1,f_2)$.  Note
that all logarithms disappear when $h_{11}$ is set to zero.

The mass, angular momentum and charge may be extracted from the
asymptotic behavior of the soliton.  In terms of the five parameters
given above, we find
\begin{eqnarray}
M&=&\ft14(h_1-3f_1+2f_2)-\ft14h_{11}(5+\ft{13}6h_1-3f_1
-h_{11}(\ft{73}{12}+h_1-2h_{11})-\ft13h_{11}\log x),\nn\\
gJ&=&\ft14(-f_1+f_2)-\ft18h_{11}(-3f_1+(2-h_{11})(2+h_1-2h_{11})),\nn\\
Q_1&=&\ft14(h_1-f_1)-\ft14h_{11}(1+\ft12h_1-\ft12h_{11}).
\label{eq:asycharges}
\end{eqnarray}
Note the $\log x$ term in the expression for the mass, which arises
because the mass is obtained from the asymptotic form of the metric in
the limit $x\to\infty$.  Clearly this indicates that the mass is
divergent, except in the case $h_{11}=0$.  This divergence also shows
up in the modified BPS expression
\begin{equation}
M=2gJ+Q_1+\ft1{12}h_{11}(h_1-\ft54h_{11}+h_{11}\log x).
\end{equation}
While at first sight this divergence may appear surprising, there is in
fact a natural explanation for where it arises.  At infinity, constant
time slices of the five-dimensional metric take the form
\begin{eqnarray}
ds^2&=&H^{1/3}\left(\fft{dx^2}{4xh}+\fft{x}4\Bigl(\sigma_1^2+\sigma_2^2
+\Bigl(h-\fft{4w_3^2}{xH}\Bigr)\sigma_3^2\Bigr)\right)\nn\\
&\sim&\fft{dx^2}{4g^2x^2}+\fft{x}4(\sigma_1^2+\sigma_2^2+(1+\ft12h_{11})
\sigma_3^2).
\end{eqnarray}
As a result, $h_{11}$ parameterizes the distortion of the $S^3$ at
infinity.  The reason for the divergent mass is simply that the space
is no longer asymptotically AdS$_5$ whenever $h_{11}\ne0$.  This squashing
of the $S^3$ is generated by a constant magnetic field at infinity
\begin{equation}
F^1\sim -\fft{h_{11}}{2g}\sigma_1\wedge\sigma_2.
\end{equation}

Turning now to the origin, demanding regularity of the soliton at
$x=0$ yields a two-parameter family of solutions specified by
$(h_0, \zeta_0)$:
\begin{eqnarray}
H&=&h_0+\fft{g^2x}2(1-h_0^2-6\zeta_0(h_0+3\zeta_0))+\cdots,\nn\\
\zeta&=&\zeta_0x-\fft{g^2x^2}2\zeta_0(h_1+3\zeta_0)+\cdots.
\end{eqnarray}
Note that the previous relation $\cosh\varphi_1=(xH_1)'$ is recovered
in the limit $\zeta_0=0$.  The matching of this expansion at $x=0$ to
the asymptotic one (\ref{eq:asyexp}) appears nontrivial but can
nevertheless be approached numerically.  We find that $h_{11}$
vanishes only when $\zeta_0=0$.  Moreover, the angular momentum $J$ in
(\ref{eq:asycharges}) also vanishes only when $\zeta_0=0$.

Turning off $\zeta_0$ yields the regular one-charge bubble of
\cite{popebubble}, with function $H_1$ given by (\ref{singleH}) (and
with $c=1$).  In general, $h_0$ is related to the $R$-charge, while
$\zeta_0$ is related to the rotation.  In this one-charge case,
non-zero rotation generates a magnetic field at infinity, resulting in
a squashing of $S^3$ and hence a divergent mass expression.  We have
also examined the three-equal charge soliton, where we found similar
behavior, except that the mass expression remains finite, even with
the magnetic field and squashing (parameterized by the analog of
$h_{11}$) present.

       The second type of solitonic solutions can be referred to as
$\R^2$ solitons, for which $0<x_0\le x<\infty$. In this case, the
geometry at $x=x_0$ is a timelike bundle over $\R^2\times S^2$.  As in
the first type of soliton, we can perform a Taylor expansion around
$x=x_0$, for which $H_I$ and $\varphi_I$ are constants at the zeroth
order, and $h(x)$, $w_3(x)$ and $U_3^I(x)$ vanish linearly when $x$
approaches $x_0$.  We used numerical methods to demonstrate that, for
appropriately chosen parameters, there are solutions for which $x$
runs smoothly from $x_0$ to $\infty$, with no CTC's.

\section{Bubble generalizations of  Klemm-Sabra solutions}

Since we have focused on an $\mathcal N=2$ truncation of the full
$\mathcal N=8$ theory, some care must be taken when counting the total
number of preserved supersymmetries.  From an $\mathcal N=8$
perspective, the general BPS bound has the form
\begin{equation}
M\ge\pm gJ_1\pm gJ_2\pm Q_1\pm Q_2\pm Q_3,
\end{equation}
where an even number of minus signs are to be taken.  For two generic
angular momenta and three generic charges, saturation of this bound
holds for only a single choice of signs.  Thus, generic three-charge
solutions with two independent rotations preserve two real
supersymmetries out of 32 ({\it i.e.,} they are 1/16 BPS in $\mathcal
N=8$).

The $\mathcal N=2$ truncation that we have taken in
Section~\ref{sec:trunc}, with $\mathcal N=2$ graviphoton given by
(\ref{eq:gphdef}), yields a BPS bound with correlated signs for the
$R$-charges:
\begin{equation}
M\ge \pm gJ_1\pm gJ_2\pm(Q_1+Q_2+Q_3).
\end{equation}
(Again, we take an even number of minus signs.)  Generic rotating
black holes then preserve 1/4 of the $\mathcal N=2$ supersymmetries,
or two real supersymmetries out of 8, in agreement with the $\mathcal
N=8$ analysis.

By focusing on a cohomogeneity one base with bi-axial symmetry, we
have essentially set the two angular momenta $J_1$ and $J_2$ equal to
each other ($J_1=J_2=J$).  In this case, the reduced BPS condition
becomes
\begin{equation}
M\ge\begin{cases}\hphantom{-}2gJ+Q_1+Q_2+Q_3\\
-2gJ+Q_1+Q_2+Q_3\\
\hphantom{-2gJ}-Q_1-Q_2-Q_3\\
\hphantom{-2gJ}-Q_1-Q_2-Q_3.\end{cases}
\label{eq:bpscoho1}
\end{equation}
The solutions that we have examined above saturate the first line of
the BPS bound, as can be seen from (\ref{BPScondition}) for the family
of solutions without hypermatter scalars.  These solutions generically
preserve 1/4 of the $\mathcal N=2$ supersymmetries, as was explicitly
demonstrated by constructing the projection (\ref{eq:14bpsproj}) out
of the hyperino variations.

It should be noted that saturation of the BPS conditions
(\ref{eq:bpscoho1}) may also be achieved by taking
\begin{equation}
M=-(Q_1+Q_2+Q_3).
\end{equation}
This gives rise to a second independent class of solutions preserving
1/2 of the $\mathcal N=2$ supersymmetries.  In fact, this family of
solutions was originally constructed by Klemm and Sabra in
\cite{Klemm:2000vn,klemmsabra} using a variety of methods including
formal analytic continuation.  Furthermore, we find that this can
be generalized by turning on the hypermatter scalars $\varphi_I$.  The
result is given by
\bea
ds^2&=&-(H_1H_2H_3)^{-2/3}(dt+w_3\sigma_3)^2+(H_1H_2H_3)^{1/3}
\left(\fft{dx^2}{4xh}+\fft{x}4(\sigma_1^2+\sigma_2^2+h\sigma_3^2)\right),
\nn\\
A^I&=&\fft1{H_I}(dt+w_3\sigma_3)-U_3^I\sigma_3,\qquad
X^I=(H_1H_2H_3)^{1/3}/H_I,\qquad\cosh\varphi_I=(xH_I)',
\label{eq:ksgen}
\eea
where
\bea
w_3 &=&\fft{1}{2} \left( g x\, H_1H_2H_3-\fft{\alpha}{x}\right),\nn\\
h &=& 1 - \fft{2g\,\alpha}{x} + g^2 x\, H_1H_2H_3,\nn\\
U_3^I&=&\fft{g}{2}x\,H_JH_K,
\label{eq:ksfuncs}
\eea
and where the functions $H_I$ obey the equation (\ref{Heqn}). As before,
the single-charge case has an explicit solution given by (\ref{singleH}).

While this Klemm-Sabra generalization is written in a similar form to
that implied by the supersymmetry analysis of
Section~\ref{sec:susyanal}, it has an important difference in that the
sign of the gauge potential in (\ref{eq:ksgen}) is {\it opposite} to
that of (\ref{eq:gaugepot}).  This suggests that the Klemm-Sabra
solution does not fall into the same class as those satisfying the
supersymmetry construction of Section~\ref{sec:susysols}, a situation
which was already hinted at in \cite{Gauntlett:2003fk}.  In fact, it
is easy to verify that the Klemm-Sabra functions (\ref{eq:ksfuncs}) do
not satisfy the relevant set of first-order equations
(\ref{firstordereqns}) found above, thus explicitly demonstrating the
incompatibility of the Klemm-Sabra solution with the construction of
Section~\ref{sec:susysols}.

Although this incompatibility might appear to demonstrate a flaw in
the supersymmetry analysis of Section~\ref{sec:susyanal} (which
purports to capture {\it all} supersymmetric solutions), this is
actually not the case.  The reason for this is that the Klemm-Sabra
family of 1/2 BPS solutions saturates the last two lines of the BPS
bound in (\ref{eq:bpscoho1}), in contrast to the 1/4 BPS solutions
which instead saturate the first.  As a result, the Klemm-Sabra
Killing spinors have a different nature from the ones constructed
above in Section~\ref{sec:hypers}.  With a different Killing spinor,
the invariant tensors (\ref{eq:bilins}) are modified, and in
particular the preferred Killing vector $K^\mu=\fft12
\overline\epsilon^i\gamma^\mu\epsilon_i$ is no longer of the form
$\partial/\partial t$ for the Klemm-Sabra solution given here.  This
indicates that, while the spatial slices of the metric
(\ref{eq:ksgen}) have the same cohomogeneity-one form as
(\ref{eq:basemet}), this metric is not the preferred K\"ahler metric
$h_{mn}$ of the base given in (\ref{eq:metspec}).  Essentially, the
Klemm-Sabra solution as written here has not been put into the
preferred coordinate system implied by (\ref{eq:metspec}), despite the
superficial similarities.

This difference in Killing spinors can be demonstrated more explicitly
by first considering the maximally symmetric AdS$_5$ vacuum written as
\begin{equation}
ds^2=-(dt+\ft12gx\sigma_3)^2+\fft{dx^2}{4xh}+\fft{x}4(\sigma_1^2
+\sigma_2^2+h\sigma_3^2),
\label{eq:adsmet}
\end{equation}
where $h=1+g^2x$.  Noting that spinors on AdS$_5$ transform as
$(\mathbf2,\mathbf1)+(\mathbf1,\mathbf2)$ under $SU(2)_L\times SU(2)_R$,
we find that the Killing spinors corresponding to the gravitino variation
of (\ref{eq:susytrans}) decompose as $\mathbf1+\mathbf1+\mathbf2$ under
$SU(2)_R$.  To see
this explicitly, it is helpful to adopt a complex spinor notation, in
which any symplectic Majorana spinor pair, say $\psi_i$, is regrouped
as a complex spinor $\psi\equiv\psi_1+i\psi_2$.  In this case, using
the vierbein basis of (\ref{eq:vierbein}) along with
$e^0=dt+\fft12gx\sigma_3$, we introduce constant complex spinors
$\chi_0^{\pm\pm}$ satisfying the mutually commuting projections
\begin{equation}
\gamma^{23}\chi_0^{\pm\alpha}=\mp i\chi_0^{\pm\alpha},\qquad
\gamma^{14}\chi_0^{\alpha\pm}=\pm i\chi_0^{\alpha\pm},
\end{equation}
where $\alpha=\pm$. Using the convention that $\gamma^{01234}=i$, the
above projections are compatible with
\begin{equation}
i\gamma^0\chi_0^{++}=-\chi_0^{++},\qquad
i\gamma^0\chi_0^{--}=-\chi_0^{--},
\label{eq:g0minus}
\end{equation}
as well as
\begin{equation}
i\gamma^0\chi_0^{+-}=\chi_0^{+-},\qquad
i\gamma^0\chi_0^{-+}=\chi_0^{-+}.
\end{equation}
Note that (\ref{eq:g0minus}) is compatible with the projection found
above in (\ref{eq:12bpsproj}).  The Killing spinors of AdS$_5$ are then
comprised of the two singlets
\begin{eqnarray}
\epsilon^{(\mathbf1)}&=&e^{\fft32igt}\chi_0^{++},\nn\\
\epsilon^{(\mathbf1')}&=&e^{-\fft12igt}[\sqrt{g^2x}-\gamma^1\sqrt{h}]
\chi_0^{-+},
\label{eq:singks}
\end{eqnarray}
as well as the doublet
\begin{equation}
\epsilon^{(\mathbf2)}=e^{-\fft12 igt}
\left(\gamma^3u+[\sqrt{g^2x}-\gamma^1\sqrt{h}]v\right)\chi_0^{+-}.
\end{equation}
The functions $u$ and $v$ are given on the $SU(2)$ orbits by
\begin{equation}
\begin{pmatrix}u\\ v\end{pmatrix}=U^{-1}\begin{pmatrix}u_0\\v_0\end{pmatrix},
\label{uveqn}
\end{equation}
where $u_0$ and $v_0$ are arbitrary constants and the $SU(2)$ matrix $U$
parameterizes the orbits.  In terms of Euler angles $(\theta,\phi,\psi)$,
$U$ may be written as
\begin{equation}
U=e^{J_3\phi}e^{J_2\theta}e^{J_3(\psi+\fft12\pi)},
\end{equation}
where $J_i=-\fft{i}2\tau_i$ with $\tau_i$ being the standard Pauli matrices.
The left-invariant one-forms $\sigma_i$ which show up in (\ref{eq:adsmet})
are given by $U^{-1}dU=J_i\sigma_i$.

Before turning on rotation, we consider the stationary BPS superstar
configuration of \cite{Behrndt:1998ns,Behrndt:1998jd}.  The canonical
form of this solution may be obtained by setting $\gamma_I=0$ in
(\ref{soln1a}).  In this case, the superstar field configuration takes the
form
\begin{eqnarray}
ds^2&=&-\mathcal H^{-2/3}(dt+w_3\sigma_3)^2+\mathcal H^{1/3}
\left(\fft{dx^2}{4xh}+\fft{x}4(\sigma_1^2+\sigma_2^2+h\sigma_3^2)\right),
\nn\\
A^I&=&-\fft1{H_I}(dt+w_3\sigma_3)+U_3^I\sigma_3,\qquad
X^I=\mathcal H^{1/3}/H_I,
\end{eqnarray}
where
\begin{equation}
w_3=\ft12gx\mathcal H,\qquad
h=1+g^2x\mathcal H,\qquad
U_3^I=\ft12gx\mathcal H/H_I,
\end{equation}
and
\begin{equation}
H_I=1+\fft{q_I}x,\qquad\mathcal H=H_1H_2H_3.
\end{equation}
This configuration breaks half of the supersymmetries,
and the surviving Killing spinors have the form
\begin{eqnarray}
\epsilon^{(\mathbf1)}&=&\mathcal H^{-1/6}\chi_0^{++},\nn\\
\epsilon^{(\mathbf1')}&=&\mathcal H^{-1/6}e^{-2igt}[\sqrt{g^2x\mathcal H}
-\gamma^1\sqrt{h}]\chi_0^{-+}.
\end{eqnarray}
These Killing spinors clearly generalize the singlet AdS$_5$ Killing
spinors (\ref{eq:singks}).  (Note that the modified time dependence is
related to the turning on of a constant gauge potential at infinity,
$A^I(\infty)=-dt$.)  After rotation is included to obtain
Gutowksi-Reall black holes \cite{Gutowski:2004ez,Gutowski:2004yv},
then only $\epsilon^{(\mathbf1)}$ survives as a Killing spinor, in
agreement with the projections (\ref{eq:12bpsproj}) and
(\ref{eq:14bpsproj}) found above.

Taking $\epsilon^{(\mathbf 1)}$ as the preferred Killing spinor, we may
verify that the $Sp(2)$ singlet bilinears take on the form
\begin{eqnarray}
f&=&\ft{i}2\overline\epsilon^{(\mathbf1)}\epsilon^{(\mathbf1)}
=\mathcal H^{-1/3},\nn\\
K^\alpha&=&\ft12\overline\epsilon^{(\mathbf1)}\gamma^\alpha\epsilon^{(\mathbf1)}
=\mathcal H^{-1/3}[1,0,0,0,0],
\end{eqnarray}
where we have normalized the constant spinor $\chi_0^{++}$ according to
$\fft{i}2\overline\chi_0^{++}\chi_0^{++}=1$.  This indicates that the
preferred Killing vector $K^\mu\partial_\mu=\partial/\partial t$ has
indeed been chosen properly to agree with the metric decomposition
chosen in (\ref{eq:metspec}).

In contrast to the Gutowksi-Reall black holes, the generalized Klemm-Sabra
solution (\ref{eq:ksgen}) with (\ref{eq:ksfuncs}) preserve the opposite
set of Killing spinors, given by the doublet
\begin{equation}
\epsilon^{(\mathbf2)}=\mathcal H^{-1/6}e^{igt}\left(\gamma^3 u+
[\sqrt{g^2 x\mathcal H}-\gamma^1\sqrt{h}]v\right)\chi_0^{+-},
\end{equation}
where $h=1+g^2x\mathcal H-2g\alpha/x$, and $u$ and $v$ are again given by 
(\ref{uveqn}).  In this case, the relevant
spinor bilinears take on a more complex form
\begin{eqnarray}
f&=&\ft{i}2\overline\epsilon^{(\mathbf2)}\epsilon^{(\mathbf2)}
=\mathcal H^{-1/3}(1-\fft{2g\alpha}x|v|^2),\nn\\
K^\alpha&=&\ft12\overline\epsilon^{(\mathbf2)}\gamma^\alpha\epsilon^{(\mathbf2)}
=\mathcal H^{-1/3}[1+(2g^2x\mathcal H-\fft{2g\alpha}x)|v|^2,0,\nn\\
&&\qquad2\sqrt{g^2x\mathcal H}\,\mathfrak{Re}(u^*v),
2\sqrt{g^2x\mathcal H}\,\mathfrak{Im}(u^*v),
2\sqrt{h}\sqrt{g^2x\mathcal H}|v|^2],
\end{eqnarray}
where now our normalization is given by $\fft{i}2\overline\chi_0^{+-}
\chi_0^{+-}=-1$ and $|u|^2+|v|^2=1$.  Although the normalization relation
$K^2=-f^2$ continues to hold (as it must by construction), this time the
preferred timelike Killing vector $K^\mu$ no longer points simply along
$\partial/\partial t$.  In fact, even in the absence of rotation ($\alpha=0$),
the Killing vector still has components along the four-dimensional base.
As a result, we see that the Klemm-Sabra solution, as given by
(\ref{eq:ksgen}), is not in the canonical form (\ref{eq:metspec}) for
the supersymmetry analysis, despite superficial appearances.

Of course, it is possible to perform an appropriate coordinate
transformation to put the Klemm-Sabra solution into canonical form.
Doing so, however, will break manifest $SU(2)_L\times U(1)$ invariance
to $U(1)^2$, corresponding to introducing a cohomogeneity-two base
appropriate to the turning on of two independent rotations.  The
Klemm-Sabra solution is then recovered in the limit when $J_1=J_2$ (in
which case the full $SU(2)_L\times U(1)$ isometry as well as
cohomogeneity-one gets restored as a hidden symmetry).

Finally, we wish to make the observation that the Killing spinors
$\epsilon^{(\mathbf1)}$, $\epsilon^{(\mathbf1')}$ and $\epsilon^{(\mathbf2)}$
(if they exist) may be put in one-to-one correspondence with the first,
second and last two lines of the BPS inequalities given in
(\ref{eq:bpscoho1}).

\section{Conclusions}

We have used the $G$-structure approach to construct supersymmetric
solutions of five-dimensional $\mathcal N=2$ gauged supergravity
coupled to two vector multiples and three incomplete hypermultiplets,
which arises from a truncation of five-dimensional $\mathcal N=8$
gauged supergravity. Different types of previously-known
supersymmetric solutions arise within this unified framework,
including rotating black holes, AdS bubbles, solitons and time
machines. New families of rotating AdS bubbles and solitonic
solutions are presented.

In addition, there are some rather exotic solitons without
hyperscalars. These include `texture'-like zero mass solitons and
`phantom'-like negative-mass solitons, where the mass is defined
relative to the AdS vacuum. These constitute
explicit examples of negative mass supergravity solutions which are
completely regular and free of closed timelike curves. In addition,
being supersymmetric guarantees that they are perturbatively free of local
instabilities.  These solutions evade positive mass theorems for
asymptotically AdS solutions by being asymptotically AdS$_5/\Z_k$ with
$k\ge 3$. In particular, the $S^3$ within AdS$_5$ has been replaced by
the lens space $S^3/\Z_k$.

It would be interesting to investigate how these exotic solitons can
be interpreted in terms of the AdS/CFT correspondence. In particular,
what is the physical quantity in the dual field theory that
corresponds to the negative mass? Determining whether the field theory
undergoes runaway behavior could offer some insight into the physical
nature of these solitons.

One could also see if asymptotically locally flat or de Sitter
solitons with negative mass could be constructed in four-dimensional
theories with zero or positive cosmological constant, and whether they
share some of the properties of the solitons discussed in this
paper. In particular, the negative mass solitons with three equal
charges do not have any scalar fields turned on. Thus, analogous
solutions might exist in four-dimensional Einstein-Maxwell de Sitter
gravity. Due to their repulsive behavior, such solitons might be used
to model the observed acceleration of the scale factor $a(t)$ of the
universe \cite{phantom1,phantom2}.

The construction of supersymmetric solutions relies upon a choice of a
four-dimensional K\"ahler base. We have limited ourselves to a
cohomogeneity-one base with bi-axial symmetry, which preserves
$SU(2)_L\times U(1)\subset SU(2)_L\times SU(2)_R\simeq SO(4)$
isometry.  This case encompasses all known black holes and AdS bubbles
with two equal rotations turned on. The more general case of two
unequal rotations would require a cohomogeneity-two base. On a similar
note, one could also consider a tri-axial four-dimensional K\"{a}hler
base space. The corresponding system of equations is presented in
Appendix B, though no solutions are known except for a couple of
special cases.

We have only considered the first-order equations for supersymmetric
backgrounds that preserve a time-like Killing vector. One could also
consider supersymmetric systems with a null Killing vector. An example
of such a solution in which the hyperscalars have not been turned on
is the magnetic string of \cite{sabra}, which was shown in
\cite{Gauntlett:2003fk} to preserve a null Killing vector.

Lastly, there are a number of other possible generalizations of the
solutions discussed in this paper, such as analogous constructions in
different dimensions as well as non-supersymmetric
generalizations, {\it e.g.,} non-extremal rotating black holes
in gauged supergravities \cite{clp1,clp2,chong1,cclp,chong2}.
Non-extremal static AdS bubbles were explored
in \cite{nonsusybubbles}. It would be interesting to investigate
whether there is a non-extremal generalization that includes both the
rotating black hole as well the rotating AdS bubble.

\section*{Acknowledgments}

We would like to thank Wei Chen and Mirjam Cveti\v{c} for helpful
conversations.  The research of H.L., C.N.P. and J.F.V.P. is supported
in part by DOE grant DE-FG03-95ER40917. The research of J.T.L. is
supported in part by DOE grant DE-FG02-95ER40899, and also by
the George P.\ and Cynthia W.\ Mitchell Institute for Fundamental
Physics.

\appendix

\section{Differential identities for the spinor bilinears}
\label{sec:diffid}

In this appendix, we present the differential identities arising from
the gravitino, hyperino and gaugino transformations.  We first present
the raw identities, and then rewrite them in a more suggestive form
notation.

The differential identities arising from the gravitino variation are
\begin{eqnarray}
\partial_\mu f&=&\ft13i_K\mathcal F,\nonumber\\
\nabla_\mu K_\nu&=&\ft13f\mathcal F_{\mu\nu}
+\ft1{12}\epsilon_{\mu\nu}{}^{\rho\lambda\sigma}\mathcal F_{\rho\lambda}
K_\sigma-\ft13W\Phi_{\mu\nu}^2,\nonumber\\
\nabla_\mu \Phi_{\nu\lambda}^a&=&\ft16(-g_{\mu[\nu}\mathcal F^{\alpha\beta}
*\Phi^a_{\lambda]\alpha\beta}+2\mathcal F_{[\nu}{}^\alpha
*\Phi^a_{\lambda]\mu\alpha}
-2\mathcal F_\mu{}^\alpha*\Phi^a_{\nu\lambda\alpha})\nonumber\\
&&-\ft23W\delta^{a2}g_{\mu[\nu}K_{\lambda]}
-\epsilon^{2ab}(g\mathcal A_\mu \Phi^b_{\nu\lambda}-
\ft13W*\Phi^b_{\mu\nu\lambda}).
\label{eq:gravids}
\end{eqnarray}
Note that the gauging explicitly breaks the $Sp(2)$ symmetry.

The hyperino transformation $\delta\lambda_{I\,i}$ from (\ref{eq:susytrans})
gives rise to
\begin{eqnarray}
i_Kd\varphi_I&=&0,\nonumber\\
f\partial_\mu\varphi_I&=&-2g\sinh\varphi_I\Phi_{\mu\nu}^2A_\nu^I,\nonumber\\
2K_{[\nu}\partial_{\mu]}\varphi_I&=&2g\sinh\varphi_I(A_\lambda^I*
\Phi^2_{\mu\nu\lambda}+X^I\Phi_{\mu\nu}^2),\nonumber\\
{}*\Phi_{\mu\nu\lambda}^a\partial_\lambda\varphi_I&=&2g\sinh\varphi_I
[\epsilon^{2ab}(*\Phi_{\mu\nu\lambda}^bA_\lambda^I+X^I\Phi_{\mu\nu}^b)
-2\delta^{a2}A_{[\mu}^IK_{\nu]}],\nonumber\\
\Phi_{\mu\nu}^a\partial_\nu\varphi_I&=&2g\sinh\varphi_I[\epsilon^{2ab}
\Phi^b_{\mu\nu}A_\nu^I+\delta^{a2}(fA_\mu^I-K_\mu X^I)],\nonumber\\
g\sinh\varphi_I(i_KA^I+fX^I)&=&0.
\end{eqnarray}
Note that, provided $g\sinh\varphi_I\ne0$, we have the condition
\begin{equation}
i_KA^I=-fX^I,
\label{eq:aicond}
\end{equation}
relating the electric potential to the scalars.  (Recall that gauge invariance
is lost when $g\varphi_I$ is turned on.)  Even for $g\sinh\varphi_I=0$,
we may take this as a gauge condition.

From the gaugino transformations $\delta\chi_i^{(\alpha)}$, we have
\begin{eqnarray}
i_Kd\phi^{(\alpha)}&=&0,\nonumber\\
fd\phi^{(\alpha)}&=&i_KF^{(\alpha)},\nonumber\\
2K_{[\nu}\partial_{\mu]}\phi^{(\alpha)}&=&[fF_{\mu\nu}^{(\alpha)}
-(i_K*F^{(\alpha)})_{\mu\nu}]+2g\partial_\alpha \hat 
W\Phi_{\mu\nu}^2,\nonumber\\
{}*\Phi^a_{\mu\nu\lambda}\partial^\lambda\phi^{(\alpha)}&=&
-2F^{(\alpha)}_{[\mu}{}^\lambda\Phi_{\nu]\lambda}^a+g\epsilon^{2ab}
\partial_\alpha \hat W\Phi_{\mu\nu}^b,\nonumber\\
\Phi_{\mu\nu}^a\partial^\nu\phi^{(\alpha)}&=&-\ft14
\epsilon_{\mu\nu\lambda\rho\sigma}F_{\nu\lambda}^{(\alpha)}\Phi_{\rho\sigma}^a
-2g\delta^{a2}\partial_\alpha\hat WK_\mu,\nonumber\\
\Phi_{\mu\nu}^aF^{(\alpha)\,\mu\nu}&=&-4g\delta^{a2}f\partial_\alpha \hat W,
\end{eqnarray}
where
\begin{eqnarray}
\phi^{(1)}&=&3\log(X^1),\nonumber\\
F_{\mu\nu}^{(1)}&=&\fft2{X^1}F_{\mu\nu}^1-\fft1{X^2}F_{\mu\nu}^2
-\fft1{X^3}F_{\mu\nu}^3,\nonumber\\
\partial_1\hat W&=&2X^1\cosh\varphi_1-X^2\cosh\varphi_2-X^3\cosh\varphi_3,
\end{eqnarray}
and similarly for $\alpha=2$.

The above identities can be put into form notation.
The gaugino and gravitino differential identities combine nicely to yield
the 0-form identities
\begin{eqnarray}
i_KdX^I&=&0,\nonumber\\
\Phi^{a\, \mu\nu}\left(\fft2{X^1}F_{\mu\nu}^1-\fft1{X^2}F_{\mu\nu}^2
-\fft1{X^3}F_{\mu\nu}^3\right)&=&-4g\delta^{a2}f(2X^1\cosh\varphi_1
-X^2\cosh\varphi_2-X^3\cosh\varphi_3),\nonumber\\
\Phi^{a\, \mu\nu}\left(-\fft1{X^1}F_{\mu\nu}^1+\fft2{X^2}F_{\mu\nu}^2
-\fft1{X^3}F_{\mu\nu}^3\right)&=&-4g\delta^{a2}f(-X^1\cosh\varphi_1
+2X^2\cosh\varphi_2-X^3\cosh\varphi_3),\nn\\
\label{eq:0formid}
\end{eqnarray}
the 1-form identities
\begin{equation}
d(fX^I)=i_KF^I,
\label{eq:1formid}
\end{equation}
the 2-form identities
\begin{equation}
d\left(\fft1{X^I}K\right)=i_K*\left(\fft1{(X^I)^2}F^I\right)+f(X^JF^K+X^KF^J)
-2g\Phi^2\cosh\varphi_I,
\label{eq:2formid}
\end{equation}
(where $I\ne J\ne K$) the 3-form identities
\begin{eqnarray}
(d\delta^{ab}+g\epsilon^{2ab}\mathcal A\wedge)\Phi^b&=&\epsilon^{2ab}
W*\Phi^b,\nonumber\\
\Phi^a\wedge d\log X^1&=&-\ft13F_\mu^{(1)\,\sigma}*\Phi^a_{\nu\lambda\sigma}
\ft12dx^\mu\wedge dx^\nu\wedge dx^\lambda-\ft13g\epsilon^{2ab}\partial_1\hat W
*\Phi^b,\nonumber\\
\Phi^a\wedge d\log X^2&=&-\ft13F_\mu^{(2)\,\sigma}*\Phi^a_{\nu\lambda\sigma}
\ft12dx^\mu\wedge dx^\nu\wedge dx^\lambda-\ft13g\epsilon^{2ab}\partial_2\hat W
*\Phi^b,
\label{eq:3formid}
\end{eqnarray}
and the 4-form identities
\begin{equation}
(d\delta^{ab}+g\epsilon^{2ab}\mathcal A\wedge)*(X^I\Phi^b)=
F^I\wedge\Phi^a+2g\delta^{a2}\left(\fft1{X^J}\cosh\varphi_K+\fft1{X^K}
\cosh\varphi_J\right)*K.
\label{eq:4formid}
\end{equation}

The hyperinos add the following:
\begin{eqnarray}
i_Kd\varphi_I&=&0,\nonumber\\
i_KA^I&=&-fX^I,\nonumber\\
fd\varphi_I&=&-2g\sinh\varphi_I\Phi^2_{\mu\nu}A^{I\, \nu} dx^\mu,\nonumber\\
d\varphi_I\wedge K&=&2g\sinh\varphi_I[X^I\Phi^2+*\Phi^2_{\mu\nu\lambda}
A^{I\,\lambda}\ft12dx^\mu\wedge dx^\nu],\nonumber\\
\Phi^a\wedge d\varphi_I&=&2g\sinh\varphi_I[\epsilon^{2ab}(\Phi^b\wedge A^I
-X^I*\Phi^b)+\delta^{a2}*(A^I\wedge K)],\nonumber\\
{}*\Phi^a\wedge d\varphi_I&=&2g\sinh\varphi_I[\epsilon^{2ab}*\Phi^b\wedge A^I
-\delta^{a2}*(fA^I-KX^I)].
\label{eq:hyperid}
\end{eqnarray}
%

\section{The tri-axial case}
\label{sec:triaxial}

A more general tri-axial class of cohomogeneity-one solutions with 
$S^3$ orbits has the following ansatz for the four-dimensional 
K\"{a}hler base:
\be
ds_4^2 = h^2\, dx^2 + \sum_{i=1}^3 a_i^2 \sigma_i^2\,,
\ee
where the functions $h$ and $a_i$ depend on $x$ only, and
the $\sigma_i$ are left-invariant one-forms satisfying
$d\sigma_i=-\fft12\epsilon^{ijk}\sigma_j\wedge\sigma_k$.
We also introduce a natural vielbein basis
\be
e^0=h\, dx\,,\qquad e^i=a_i\, \sigma_i\,.
\ee
An ansatz for the $SU(2)$ invariant anti-self-dual K\"{a}hler form is
\be
J=\sum_{i=1}^3 \alpha_i(x) (e^0\wedge e^i-\ft12\epsilon_{ijk} e^j\wedge e^k)\,,
\ee
where $\sum_i \alpha_i^2=1$. The base metric is K\"{a}hler if $J$ is
covariantly constant, which implies that
\be
\alpha_1'=\Big( \fft{h(a_3^2-a_1^2-a_2^2)}{2a_1a_2a_3}-
\fft{a_3'}{a_3}\Big) \alpha_2+ \Big( 
\fft{a_2'}{a_2}-\fft{h(a_2^2-a_1^2-a_3^2)}{2a_1a_2a_3}\Big) 
\alpha_3\,,\qquad {\rm and\ cyclic}\,,
\ee
where a prime denotes a derivative with respect to $x$. For
simplicity, we will take $\alpha_1=\alpha_2=0$ and $\alpha_3=1$, so
that
\be
J=e^0\wedge e^3 - e^1\wedge e^2\,,
\ee
and
\bea
2a_1' a_2a_3 &=& h(a_2^2+a_3^2-a_1^2)\,,\nn\\
2a_1a_2'a_3 &=& h(a_1^2+a_3^2-a_2^2)\,.\label{kahlercond}
\eea
Note that these two above conditions imply that
\be
h\, a_3=(a_1\,a_2)'\,.
\ee
The base also has the Ricci form
\bea
\mathcal R &=& 
\fft{1}{2a_1a_2a_3} \Big[ \fft{a_3^4-(a_1^2-a_2^2)^2}{a_1a_2a_3}
-\fft{2 a_1 a_2 a_3''}{h} + \fft{2 a_1 a_2 a_3' h'}{h^3}
  - \fft{2 a_3 a_3'}{h}\Big]\, e^0\wedge e^3\nn\\
&&
+\fft{2a_1a_2a_3'+(a_3^2-a_1^2-a_2^2)h}{2a_1^2a_2^2h}\,e^1\wedge e^2\,,
\eea
which can be expressed as
\be
\mathcal R=d\Big[ \Big( \fft{a_1^2+a_2^2- a_3^2}{2a_1a_2}-
\fft{a_3'}{h}\Big)\,\sigma_3\Big]\,.
\ee

We make the following ansatz for the one-forms $\omega$ and $\beta^I$:
\begin{equation}
\omega=\sum_{i=1}^3 w_i\sigma_i,\qquad\beta^I=\sum_{i=1}^3 U_i^I \sigma_i.
\end{equation}
$d\omega$ decomposes into self-dual and anti-self dual components according to
\be
(d\omega)^{\pm}=\ft12 \sum_{i\ne j\ne k} \Big( \fft{w_i'}{ha_i}\mp
\fft{w_i}{a_ja_k}\Big) (e^0\wedge e^i\pm \ft12 \epsilon^{ijk} e^j\wedge e^k)\,.
\ee
$(d\beta^I)^\pm$ have the same form as $(d\omega)^{\pm}$, except with
$w_i\rightarrow U_i^I$.

Inserting these expressions into the supersymmetry conditions
(\ref{eq:finsusycond}) gives rise to the first-order equations
\begin{eqnarray}
\varphi_I'&=&-2g \fft{h U_3^I}{a_3} \sinh\varphi_I,\nn\\
\fft{U_3^{I'}}{ha_3}+\fft{U_3^I}{a_1a_2} &=& 2g(H_J
\cosh\varphi_K+H_K\cosh\varphi_J),\qquad
\,\,\, I\ne J\ne K\nn\\
U_1^{I'} &=& -\fft{ha_1}{a_2a_3} U_1^I\,,\qquad U_2^{I'}=
-\fft{ha_2}{a_1a_3} U_2^I,\nn\\
w_i'-\fft{ha_i}{a_ja_k} w_i &=& \ft12 \sum_I H_I
\left( U_i^{I'}-\fft{ha_i}{a_ja_k} U_i^I\right)\,,\qquad 
\qquad i\ne j\ne k
\nn\\
\fft{a_3^2-a_1^2-a_2^2}{2a_1a_2}+\fft{a_3'}{h} &=&
g\sum_I U_3^I \cosh\varphi_I\,,\label{firstordereqns1}
\end{eqnarray}
as well as the algebraic conditions
\be
\sum_I U_{\ell}^I \cosh\varphi_I=0\,,\qquad gU_{\ell}^I\sinh\varphi_I=0\,,
\ee
where $\ell=1,2$.

In addition, the second-order equation of motion (\ref{eq:fineom})
reduces to
\begin{equation}
0=\Big( \fft{a_1a_2a_3}{h} H_I'+\sum_i U_i^JU_i^K\Big)' -
2g\cosh\varphi_I(a_1a_2w_3'+ha_3w_3) +4g^2 ha_1a_2a_3\,
\sinh^2\varphi_IxH_JH_K.\label{secondordereqn1}
\end{equation}
Consider the purely gravitational system with $\varphi_I=U_1^I=U_2^I=0$,
$H_I=1$, for which 
\bea
U_3^I &=& w_3=2ga_1a_2\,,\nn\\
w_{\ell} &=& c_{\ell} \exp\Big[ \int^x dx\, \fft{ha_3}{a_1a_2}\Big]\,.
\eea
Then the base space is Einstein-K\"{a}hler, and is described
by the functions $a_i$ which obey the equations (\ref{kahlercond})
along with
\be
2a_1a_2a_3'=(a_1^2+a_2^2-a_3^2+12g^2 a_1^2a_2^2)\,,
\ee
where we have chosen a gauge such that $h=1$. This system of equations has 
been considered by Dancer and Strachan \cite{dancer}. Explicit solutions 
are only known for comparatively simple examples such as the tri-axial 
forms of the Fubini-Study metric on $\CP^2$ and the product metric on 
$\CP^1\times \CP^1$. Since this does not bode well for finding explicit 
solutions with additional fields, in this paper we have focused on the 
bi-axial case $a_1=a_2$.


\end{document}